\documentclass[prd,nofootinbib,twocolumn,superscriptaddress,preprintnumbers,balancelastpage]{revtex4-1}

\usepackage{aas_macros}
\usepackage{amsmath,amssymb,mathtools,bm}
\usepackage{graphicx, color, hepunits}
\usepackage[dvipsnames]{xcolor}
\usepackage{float}
\usepackage{filecontents}
\usepackage{multirow}
 \usepackage{hyperref} 
\hypersetup{
    colorlinks=true,       
    linkcolor=blue,        
    citecolor=blue,        
    filecolor=magenta,     
    urlcolor=blue          
}
\usepackage[utf8]{inputenc}
\usepackage[english]{babel}
\usepackage[nolist]{acronym}
\usepackage{xfrac}
\usepackage[normalem]{ulem}

\let\vec\mathbf

\def\mysection#1{
{\it #1.---} }

\newcommand{\es}[2] {\begin{equation} \label{#1} \begin{split} #2 \end{split} \end{equation}}

\begin{document}
\preprint{MIT-CTP/5292}
\begin{acronym}
\acro{ADM}{axion dark datter}
\acro{SM}{Standard Model}
\acro{QCD}{quantum chromodynamics}
\acro{PQ}{Peccei-Quinn}
\acro{ALP}{axion-like particle}
\acro{WIMP}{Weakly Interacting Massive Particle}
\acro{PSD}{power spectral densitie}
\acro{DM}{dark matter}
\acro{DFT}{discrete Fourier transform}
\acro{FLL}{flux-lock feedback loop}
\acro{SNR}{signal-to-noise ratio}
\acro{LEE}{look-elsewhere effect}
\acro{TS}{test statistic}
\acro{POM}{polyoxymethylene}
\acro{PTFE}{polytetrafluoroethylene}
\acro{MC}{Monte Carlo}
\acro{AFS}{active feedback stabilization}
\acro{DR}{dilution refrigerator}
\end{acronym}

\newcommand{\ABRA}{\mbox{ABRACADABRA}\xspace}
\newcommand{\abra}{\mbox{ABRACADABRA-10\,cm}\xspace}
\newcommand{\SM}{\ac{SM}\xspace}
\newcommand{\PQ}{\ac{PQ}\xspace}
\newcommand{\QCD}{\ac{QCD}\xspace}
\newcommand{\ADM}{\ac{ADM}\xspace}
\newcommand{\DM}{\ac{DM}\xspace}
\newcommand{\ALP}{\ac{ALP}\xspace}
\newcommand{\ALPs}{\ac{ALP}s\xspace}
\newcommand{\WIMP}{\ac{WIMP}\xspace}

\newcommand{\AFS}{\ac{AFS}\xspace}
\newcommand{\PSD}{\ac{PSD}\xspace}
\newcommand{\PSDs}{\acp{PSD}\xspace}
\newcommand{\DFT}{\ac{DFT}\xspace}
\newcommand{\FLL}{\ac{FLL}\xspace}
\newcommand{\SNR}{\ac{SNR}\xspace}
\newcommand{\LEE}{\ac{LEE}\xspace}
\newcommand{\TS}{\ac{TS}\xspace}
\newcommand{\POM}{\ac{POM}\xspace}
\newcommand{\PTFE}{\ac{PTFE}\xspace}
\newcommand{\MC}{\ac{MC}\xspace}
\newcommand{\DR}{\ac{DR}\xspace}

\newcommand{\rhoDM}{\ensuremath{\rho_{\rm DM}}\xspace}
\newcommand{\gagg}{\ensuremath{g_{a\gamma\gamma}}\xspace}
\newcommand{\Jeff}{\ensuremath{\mathbf{J}_{\rm eff}}\xspace}

\newcommand{\Px}{\ensuremath{\bar{\mathcal{F}}}}
\newcommand{\Pten}{\ensuremath{\bar{\mathcal{F}}_\mathrm{10M}}\xspace}
\newcommand{\Pone}{\ensuremath{\bar{\mathcal{F}}_\mathrm{1M}}\xspace}
\newcommand{\Phun}{\ensuremath{\mathcal{F}_\mathrm{100k}}\xspace}

\title{The Galactic potential and dark matter density from angular stellar accelerations}
\date{\today}
\author{Malte Buschmann}
\email{msab@princeton.edu}
\affiliation{Department of Physics, Princeton University, Princeton, NJ 08544, USA}


\author{Benjamin~R.~Safdi}
\email{brsafdi@lbl.gov}
\affiliation{Berkeley Center for Theoretical Physics, University of California, Berkeley, CA 94720, USA}
\affiliation{Theoretical Physics Group, Lawrence Berkeley National Laboratory, Berkeley, CA 94720, USA}

\author{Katelin Schutz}\thanks{Einstein Fellow}
\email{kschutz@mit.edu}
\affiliation{Center for Theoretical Physics, Massachusetts Institute of Technology, Cambridge, MA 02139, USA
}

\begin{abstract}
\noindent 
We present an approach to measure the Milky Way (MW) potential using the angular accelerations of stars \emph{in aggregate} as measured by astrometric surveys like {\it Gaia}. Accelerations directly probe the gradient of the MW potential, as opposed to indirect methods using {\it e.g.} stellar velocities.
We show that end-of-mission {\it Gaia} stellar acceleration data may be used to measure the potential of the MW disk at approximately 3$\sigma$ significance and, if recent measurements of the solar acceleration are included, the local dark matter density at $\sim$2$\sigma$ significance. Since the significance of detection scales steeply as $t^{5/2}$ for observing time $t$, future surveys that include angular accelerations in the astrometric solutions may be combined with {\it Gaia} to precisely measure the local dark matter density and shape of the density profile.

\end{abstract}

\maketitle

The density distribution of matter in the Milky Way (MW) is the most fundamental quantity that can be measured in Galactic dynamics, underlying virtually every study of the assembly history and evolution of the MW and its satellites. Moreover, the local density of the dark matter (DM) component of the MW is a central input for searches for particle DM in terrestrial experiments and in indirect searches: the translation of experimental results to particle DM properties relies directly on spatial and kinematic knowledge of the DM halo in which the experiment is embedded. A robust determination of the matter distribution in the Galaxy, using methods that make as few assumptions as possible, is therefore of paramount importance. In this work we present a novel approach to this problem making use of angular stellar accelerations that may be measured in astrometric surveys. 

The matter content in the MW is often separated into morphologically distinct disk, bulge, and DM halo components. The local density of baryonic components in the MW can be inferred somewhat directly via observations of stars and measurements of the column density of certain gas components, although astrophysical modelling is needed to account for biases, degeneracies, and unobserved components (see Ref.~\cite{mckee2015stars} for a review and Ref.~\cite{2017MNRAS.470.1360B} for a post-\emph{Gaia} update of the stellar inventory). Meanwhile the distribution of DM must be deduced indirectly through its gravitational influence on stellar velocities via: (i) the use of the Galactic rotation curve (see, {\it e.g.},~\cite{2019ApJ...871..120E, Benito:2020lgu}) or (ii) the use of analyses that simultaneously describe the velocity and spatial distributions of tracer stars ({\it e.g.}, through Jeans equations)~\cite{kuijken1989mass,holmberg2000local,Bovy:2012tw,Schutz2018,widmark2019measuring}. The interpretation of the rotation curve is strongly subject to priors, since the data can only determine the mass enclosed within a given radius; additional information is required to separate that enclosed mass into bulge, disk, and DM components~\cite{deSalas:2019pee}. The Jeans-type methods, on the other hand, often involve simplifying assumptions about axisymmetry, equilibrium, and separability of motion in different directions. Evidence for out-of-equilibrium effects can manifest as vertical density and velocity waves~\cite{bennett2019vertical}, phase-space spirals~\cite{2018Natur.561..360A}, and merger remnants~\cite{helmi2018merger}, which complicates these analyses. 

Recently a more direct method has emerged for probing the DM distribution of the MW using stellar accelerations. By the Poisson equation, $\nabla^2 \Phi = -\nabla \cdot {\bf a}^{\rm GF} =  4\pi G \rho$, the Galactic-frame gravitational acceleration ${\bf a}^{\rm GF}$ may be directly related to the gravitational potential $\Phi$ and mass density $\rho$, regardless of whether the tracer stars are in equilibrium. Since this approach does not rely on taking averages or moments of kinematic distributions, it is independent of the selection function and completeness of a given survey of stellar motions. Recent work has shown that spectral measurements of the line-of-sight (radial) velocities of \emph{individual} stars will be sensitive to the local DM density~\cite{Ravi:2018vqd,Silverwood:2018qra,Chakrabarti:2020kco}. Line-of-sight accelerations of pulsars measured via their spins~\cite{Phillips:2020xmf} and orbital periods~\cite{Phillips:2020xmf,Chakrabarti:2020abx, bovy2020purely} may already have hints of the local Galactic acceleration, with improvements possible in the future.

\begin{figure}[htb]
\begin{center}
\includegraphics[width = .49\textwidth]{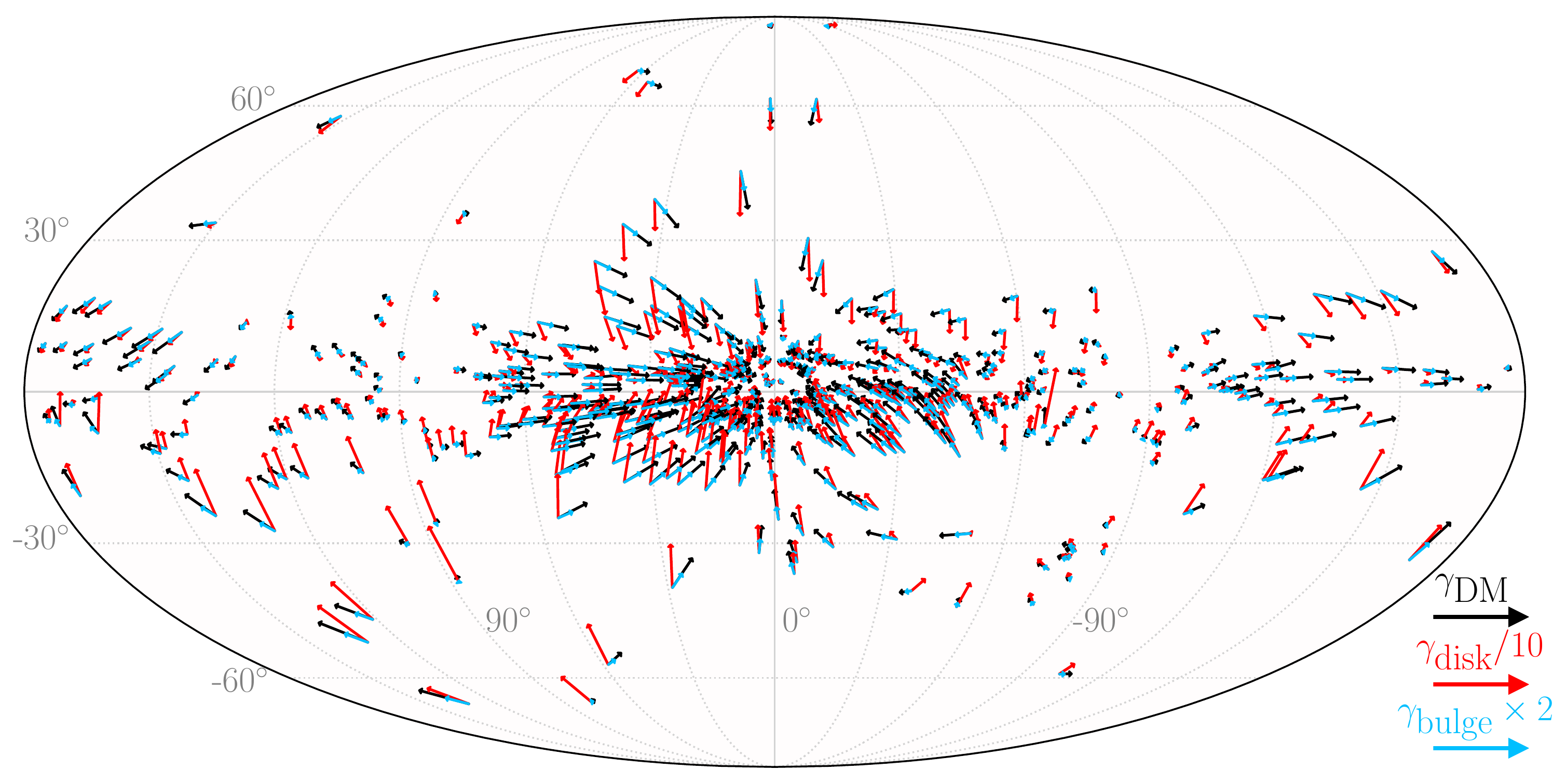}
\end{center}
\vspace{-0.4cm}
\caption{The stellar accelerations in the non-inertial solar frame, in Galactic coordinates with the Galactic Center at the center, in our fiducial MW potential model for a sample of 512 randomly selected stars with $\varpi / \sigma_{\varpi}  >66$ from the {\it Gaia} EDR3 catalog, with contributions from the bulge (blue), DM halo (black), and disk (red) separated. The arrows point in the directions of the acceleration vectors, with the lengths proportional to the magnitudes (note that the disk and bulge accelerations are rescaled compared to the DM for illustration purposes). In the Galactic frame, the bulge and halo accelerations point toward the Galactic center with magnitudes that vary differently with Galactocentric radius. The disk accelerations clearly point more towards the Galactic plane (note that in the solar frame there is nontrivial apparent acceleration away from the Galactic Centre at large longitudes). 
\vspace{-0.4cm}} 
\label{fig:accel_ill}
\end{figure}

In this work we show that angular stellar accelerations from astrometric surveys like {\it Gaia} may provide a complementary acceleration-based measure of the matter content of the MW. The expected relative acceleration of stars in the solar neighborhood is $\mathcal{O}(0.1)\,$cm/s/yr, corresponding to $\sim$nanoarcsecond/yr proper motion (PM) differences over the course of a decade for stars at $\sim$kpc distances. This minuscule level of angular acceleration from the MW potential is not detectable for an individual star. However, the \emph{combined} effect on the $N \sim 10^9$ stars observed by {\it Gaia} and similar surveys may be discerned through the use of a joint likelihood. We project that end-of-mission data from {\it Gaia} should be able to provide more than $\sim$3$\sigma$ evidence for the acceleration induced by the MW disk.
The magnitude and direction of the solar acceleration were recently determined through the apparent PM of distant quasars in {\it Gaia}~EDR3 data caused by the acceleration-induced aberration effect~\cite{2020arXiv201202036G}. We show that combining this measurement with {\it Gaia} acceleration data may allow for a measurement at the $\sim$1--2$\sigma$ level of the local DM density. The detection significance grows rapidly with time $t$ as $t^{5/2}$, meaning that future surveys may combine their data with {\it Gaia} to separately measure the DM content of the MW and the Galactic disk and bulge at high precision, even without input from other techniques. We demonstrate the feasibly of this approach by performing an analysis using acceleration data inferred from the combination of {\it Gaia} DR2 data~\cite{Brown2018,2018A&A...616A...2L} and {\it Hipparcos} data~\cite{1997ESASP1200.....E,2018ApJS..239...31B}, with results consistent with those expected from statistical uncertainties alone.

\mysection{Angular accelerations}
Typically, astrometric solutions assume five parameters to specify the apparent angular motion: the right ascension (RA) $\alpha$, declination (DEC) $\delta$, PM $\mu_\alpha$ and $\mu_\delta$ in these two directions, respectively, and the parallax $\varpi$. In this work, we advocate that two additional parameters be included in the astrometric solution: the angular acceleration in RA and DEC, which we denote by
$\gamma_\alpha$ and $\gamma_\delta$. Such astrometric solutions may be determined by fitting the positions of $\mathcal{N}$ images acquired over time with terms that are linear and quadratic in the observing time $t$, accounting for PM and angular acceleration, respectively, in addition to parallax. In the large-$\mathcal{N}$ limit and assuming $\mathcal{N} \propto t$, the uncertainty on the RA and DEC of a given star scales with time as
$\sigma_{\alpha, \delta} \propto 1/\sqrt{t}$. Since the size of the angular offset since the initial observation grows linearly and quadratically with $t$ for PM and angular acceleration, respectively, those uncertainties, which we denote by $\sigma_\mu$ and $\sigma_\gamma$, decline with additional powers of $t$. Explicitly, we expect the scaling $\sigma_\mu \sim 8 \sigma_{\alpha, \delta}/ \sqrt{3}\,t $ and $\sigma_\gamma \sim \sqrt{15}\, \sigma_\mu / 2 t$~\cite{VanTilburg:2018ykj}. Given that accelerations are not included in current astrometric solutions for {\it Gaia}, we will use these relations to {\it project} acceleration uncertainties that may be produced in future astrometric solutions with accelerations included. 

The seven-parameter astrometric solution allows us to determine the apparent acceleration of the star projected onto the celestial sphere in the non-inertial solar frame. To compute the accelerations in the inertial Galactic frame we must account for the acceleration of the Sun itself from the Galactic potential. We define the acceleration of the Sun due to the Galactic potential in the Galactic frame as ${\bf a}_\odot = - \nabla \Phi({\bf r}_\odot)$, where ${\bf r}_\odot$ denotes the solar position and the distance $r_\odot = 8.224 \pm 0.071$ kpc has recently been measured precisely by the GRAVITY collaboration~\cite{du2018detection,abuter2020arxiv,2019A&A...625L..10G,bovy2020purely}. Note that for most of this work we do not incorporate prior information on the solar acceleration but determine it self consistently from the MW potential model. In the inertial Galactic frame a given star at position ${\bf r}_i$ experiences an acceleration ${\bf a}_i^{\rm GF} = - \nabla \Phi({\bf r}_i)$, but in the non-inertial solar frame (in which observations are made) the apparent acceleration is \mbox{${\bf a}_i = {\bf a}_i^{\rm GF} - {\bf a}_\odot$}. The projections of these apparent accelerations onto the celestial sphere are illustrated in Fig.~\ref{fig:accel_ill} for a random sample of 512 {\it Gaia} EDR3 stars, with parallax uncertainties satisfying $\varpi / \sigma_{\varpi}  >66$, for our fiducial MW potential model, which is described below.

In this work, we assume that the Galactic gravitational potential is solely responsible for correlated apparent accelerations observed for a sample of $N$ stars in the solar neighborhood. Known external sources of acceleration, for instance the Large Magellanic Cloud~\cite{2019MNRAS.487.2685E,2020arXiv201013789E,2021MNRAS.501.2279V}, contribute to the signal at the percent level compared to the acceleration induced by the MW potential~\cite{2020arXiv201202036G,bovy2020purely}. Additionally, previous studies of stellar accelerations have found that planetary and binary companions contaminate the acceleration signal of \emph{individual} stars at a negligible level~\cite{Ravi:2018vqd,Chakrabarti:2020kco}, and we expect this to be especially true when looking at the aggregate acceleration since acceleration kicks by companions in random directions relative to the observer will average down. We therefore directly equate measured accelerations in this work with the gravitational potential of the MW, up to statistical uncertainties. We parameterize the potential as $\Phi = \Phi_{\rm bulge} + \Phi_{\rm disk}  + \Phi_{\rm DM}$, with contributions from the bulge, disk, and DM halo, respectively. We treat the bulge as a Hernquist sphere~\cite{1990ApJ...356..359H} with scale radius of 0.6~kpc~\cite{2013ApJ...779..115B} (noting that for most applications the stars we use to trace the acceleration will be well outside the bulge, indicating that the main parameter of interest is the total mass, which we take to be $5 \times 10^9$ $M_\odot$~\cite{2015ApJS..216...29B}). We treat the disk as a Miyamoto-Nagai disk~\cite{1975PASJ...27..533M} with a 3~kpc scale length, 280~pc scale height, and mass of $6.8 \times 10^{10}$ $M_\odot$~\cite{2015ApJS..216...29B}. We model the DM halo as a Navarro-Frenk-White (NFW) profile with scale radius $r_s=16$~kpc and a local DM density $\rho_\odot= 0.3\, \text{GeV}/\text{cm}^3 = 0.008 \, M_\odot/\text{pc}^3$~\cite{2015ApJS..216...29B} (which is broadly consistent with more recent determinations with {\it Gaia}~\cite{cautun2020milky,nitschai2020first}). Further details can be found in the Supplemental Material (SM). For the purposes of this work, we introduce normalization parameters for the disk, DM halo, and bulge contributions to the potential ($\lambda_{\rm disk}$, $\lambda_{\rm DM}$, $\lambda_{\rm bulge}$), with unity corresponding to the true value in our fiducial model. We will also consider a one-parameter model with parameter $\lambda$ that re-scales the entire gravitational potential, with fixed normalization between the three sub-components. For most of this work we also fix the morphological parameters of the potential for simplicity, such as the disk scale parameters and $r_s$, though these parameters are also measurable through the joint likelihood, as we indicate below.

\mysection{Statistical Analysis Framework} We employ a likelihood-based framework to constrain the Galactic potential with stellar acceleration data. The dominant statistical uncertainties entering into the likelihood are those on the accelerations, while the uncertainties on $\alpha$, $\delta$, and $\varpi$ play sub-dominant roles and may be neglected. Note that PMs do not enter into the likelihood. See the SM for an extended discussion of these points. 

For a given star indexed by $i$, we define the vector
\es{eq:x}{
{\bf x}_i({\bm \theta}) \equiv \begin{pmatrix} {a_{\alpha, i}({\bm \theta}, \alpha_i, \delta_i, \varpi_i) \varpi_i - \gamma_{\alpha, i} }\\ {a_{\delta, i}({\bm \theta}, \alpha_i, \delta_i, \varpi_i) \varpi_i - \gamma_{\delta, i} } \end{pmatrix} \,,
} 
where $a_{\alpha, i}({\bm \theta}, \alpha_i, \delta_i, \varpi_i)$ and $a_{\delta, i}({\bm \theta}, \alpha_i, \delta_, \varpi_i)$ are the model predictions for the apparent projected acceleration components of the star and $\{\gamma_{\alpha_i}, \gamma_{\delta_i}\}$ denote the measured angular accelerations. The probability of observing the data $\vec{d}_i = \{\alpha_i, \delta_i, \varpi_i, \gamma_{\alpha, i}, \gamma_{\delta, i} \}$ given the model parameters ${\bm \theta}$, which parameterize the potential, is
\es{eq:p_i_deff}{
p_i(\vec{d}_i| {\bm \theta}) = {1 \over \sqrt{ \det(2 \pi {\bf \Sigma}_i)}} e^{ - {1 \over 2} {{\bf x}_i \cdot {\bf \Sigma}_i^{-1} \cdot {\bf x}_i }} \,,
}
where ${\bf \Sigma}_i$ is the covariance matrix between the angular acceleration components from the astrometric solution for the star. Assuming measurements of individual stars are independent, we may then construct a joint likelihood \mbox{$p({\bf d} | {\bm \theta}) = \prod_{i=1}^N p_i({\bf d}_i | {\bm \theta})$} with ${\bf d} = \{ {\bf d}_i \}_{i=1}^N$ for $N$ stars. In this work we perform frequentist inference using the joint likelihood, for example defining the discovery test statistic \mbox{$q \equiv 2 [ \ln p({\bf d}| \hat {\bm \theta}) - \ln p({\bf d}| \hat {\bm 0}) ]$}, with $\hat {\bm \theta}$ denoting the best-fit model parameters that maximize the likelihood and ${\bm 0}$ the null model with no acceleration.

Before performing numerical sensitivity projections it is instructive to analytically estimate the parametric sensitivity of a stellar acceleration survey to the acceleration from the MW potential. For simplicity, in the following estimate we assume that all stars are in the plane of the Galaxy, assume that the Galaxy is azimuthally symmetric, and work to leading non-trivial order in $r_i / r_\odot$, with $r_i$ the distance of the star from the Sun. Later in this work we will perform numerical analyses without such assumptions. Additionally, we invoke the Asimov data set~\cite{Cowan:2010js} and take the data to be the mean expectation under the signal hypothesis. Using the Asimov data set allows us to estimate, for example, the mean expected discovery test statistic without having to perform multiple Monte Carlo (MC) realizations. The signal hypothesis is the model described by our fiducial MW potential model while the null hypothesis has no Galactic accelerations. As derived in the SM, the discovery test statistic is approximated by
\es{eq:delta_chi2_scale}{
q \approx {N \langle \sigma_{\gamma}^{-2} \rangle \over 8  }  \left(3{a_\odot \over r_\odot} + 4 \pi G \bar \rho_\odot\right)^2 \,,}
where $\bar \rho_\odot = \int d \Omega \rho(r_\odot, \Omega) / (4 \pi)$ is the galactocentric-angle-averaged matter density at the solar location. Note that even though locally the matter density is dominated by the disk component, $\bar \rho_\odot$ is dominated by DM since this is the average density in a spherical shell with radius $r_\odot$. The quantity $\langle \sigma_{\gamma}^{-2} \rangle$ is the catalog-averaged inverse variance in the angular acceleration. 

We now use~\eqref{eq:delta_chi2_scale} to perform a rough estimate of the sensitivity of {\it Gaia} to the stellar accelerations, with more rigorous numerical forecasts presented below. For each of the $N = 1,110,324,277$ stars in the {\it Gaia} EDR3 catalog~\cite{2020arXiv201202036G} with positive parallax we use the PM uncertainties and projected relation between PM uncertainties and acceleration uncertainties~\cite{VanTilburg:2018ykj} 
to project the acceleration uncertainty that would be obtained in a seven-parameter astrometric solution incorporating $t$ years of data. Note that the {\it Gaia} EDR3 catalog used $t \approx 2.76$ yrs of data. We estimate $\langle \sigma_{\gamma}^{-2} \rangle \approx (2.8 \, \, \mu{\rm as}/{\rm yr}^2)^{-2} (t / 10 \, \, {\rm yr})^5$ for the EDR3 stars with positive parallax, which leads to the predicted discovery test statistic $q \approx 7.7 \left({ t / 10 \, \, {\rm yr}}  \right)^5$, assuming $a_\odot / r_\odot \approx 9 \times 10^{-16} / {\rm yr}^2$~\cite{2020arXiv201202036G,bovy2020purely}. Invoking Wilks' theorem, we expect $q$ to follow a chi-square distribution with one degree of freedom (the parameter that rescales the full MW potential), which implies that we may interpret $\sqrt{q}$ as the detection significance. Thus, if {\it Gaia} were to take data for $\sim$10 yrs, which is roughly the possible extent of its lifetime, then we would expect from this estimate a $\sim$$3$$\sigma$ detection of the Galactic acceleration. Importantly, though, the test statistic scales rapidly with time; {\it e.g.}, taking $t \approx 20$ years would lead to over 10$\sigma$ evidence for the Galactic acceleration. 

We may also, within this simplified framework, consider the two-parameter model with model parameters $a_\odot/r_\odot$ and $\bar \rho_\odot$. While these parameters could be constrained simultaneously from the data, we may incorporate prior knowledge of {\it e.g.} $a_\odot/r_\odot$, which has recently been measured precisely by {\it Gaia} as $a_\odot/r_\odot = (9.12 \pm 0.63) \times 10^{-16}/{\rm yr}^2$~\cite{2020arXiv201202036G,bovy2020purely}. In the limit where we fix $a_\odot/r_\odot$ and only consider the model parameter $\bar \rho_\odot$, we may use~\eqref{eq:delta_chi2_scale} to infer that with $t = 10$ years of {\it Gaia} data we may measure $\rho_{\rm DM}$ at $\sim$1.4$\sigma$ precision. In the following Sections, we perform more precise sensitivity estimates that go beyond the analytic approximation in~\eqref{eq:delta_chi2_scale}, starting with an analysis of {\it Gaia-Hipparcos} data before returning to projections for future {\it Gaia} data releases.

\mysection{{\it Gaia}-{\it Hipparcos} data}In Ref.~\cite{2018ApJS..239...31B} data from the {\it Hipparcos} and {\it Gaia} DR2 catalogs were combined to produce a catalog of stellar accelerations. In the SM we describe additional post-processing that we perform in order to extract the $\gamma_\alpha$, $\gamma_\delta$, and the associated covariance matrices. Note that the {\it Hipparcos} and {\it Gaia} DR2 catalogs were separated by a baseline of $\sim$24 yrs. The combined acceleration catalog, subject to quality cuts, contains $N = 86,201$ stars with $\langle \sigma_\gamma^{-2} \rangle \approx \left( 14 \, \, \mu{\rm as}/{\rm yr}^2 \right)^{-2}$. Using~\eqref{eq:delta_chi2_scale} we thus estimate a discovery test statistic $q \sim 2 \times 10^{-5}$ for the Asimov data set, meaning that with the {\it Gaia}-{\it Hipparcos} acceleration catalog we are not able to detect the acceleration of the MW's potential but we should be able to constrain potentials of order $\sim {\rm few}\times 10^2$ larger than the true values. 

To verify our estimate, we first implement two quality cuts on the {\it Gaia}-{\it Hipparcos} catalog. We exclude stars with PM uncertainties larger than 0.7 mas/yr and outlier stars that have accelerations over 4$\sigma$ away from zero. The first cut removes about $\sim$6\% of the total stars and addresses the non-Gaussian tails in the distributions of accelerations that these stars produce, as discussed in~\cite{2018ApJS..239...31B}. The second cut reduces the number of stars by a further $\sim$20\%. We confirm in the SM that this cut does not bias the analysis, since the contribution of the MW's potential to any individual star's acceleration is negligible compared to the per-star acceleration uncertainties. 

We consider the one-parameter model with model parameter $\lambda$ that re-scales the full MW potential, with the height of the Sun fixed at $z = 0.02$ kpc~\cite{2019MNRAS.482.1417B}. 
Using the joint likelihood and the {\it Gaia}-{\it Hipparcos} data set we find a best-fit re-scaling parameter $\hat \lambda \approx 10^2$ with a discovery test statistic of $q \approx 1.5$. This is consistent with the expected normalization ($\lambda = 1$) at $\sim$1.2$\sigma$, and at 95\% confidence $\lambda \lesssim 220$.
In the SM we present an approach, in the context of the {\it Gaia}-{\it Hipparcos} analysis but that may be applied more generally, for testing and accounting for systematic uncertainties by scrambling the mapping from model predictions to stellar data.

\mysection{Simulated {\it Gaia} data} 
The {\it Gaia}-{\it Hipparcos} analysis indicates that we need to improve the sensitivity to the gravitational potential by approximately a factor of $\sim$100 in order to detect the acceleration from the MW. This increase in sensitivity, as we now show, may come with future {\it Gaia} data releases, if the accelerations are included in the astrometric solution. In particular, we perform projections using Asimov data constructed, as described previously, by re-scaling the current {\it Gaia} EDR3 PM covariance matrices to anticipate the covariance matrices associated with future astrometric solutions including accelerations. We then analyze the Asimov data using the full joint likelihood.

As with the {\it Gaia}-{\it Hipparcos} analysis, we begin by considering the 1-parameter model where the entire MW potential is re-scaled relative to the fiducial normalization by $\lambda$, with truth value $\lambda = 1$. By computing the discovery test statistic on the Asimov data set we infer $q \approx 13.2 \times (t / 10 \, \, {\rm yr} )^{5/2}$, meaning that with 10 years of {\it Gaia} data the acceleration from the potential of the MW should be detectable at the level $\sim$3.6$\sigma$. This estimate of $q$ is similar to but slightly larger than that we obtained using~\eqref{eq:delta_chi2_scale}, which is partially the result of the approximation in~\eqref{eq:delta_chi2_scale} not accounting for acceleration components towards the Galactic plane.

Next we analyze the multi-parameter model with ${\bm \theta} = ( \lambda_{\rm disk}, \lambda_{\rm bulge}, \lambda_{\rm DM})$ that allows us to re-scale the three different potential components independently. 
\begin{figure}[t!]
\begin{center}
\includegraphics[width = .49\textwidth]{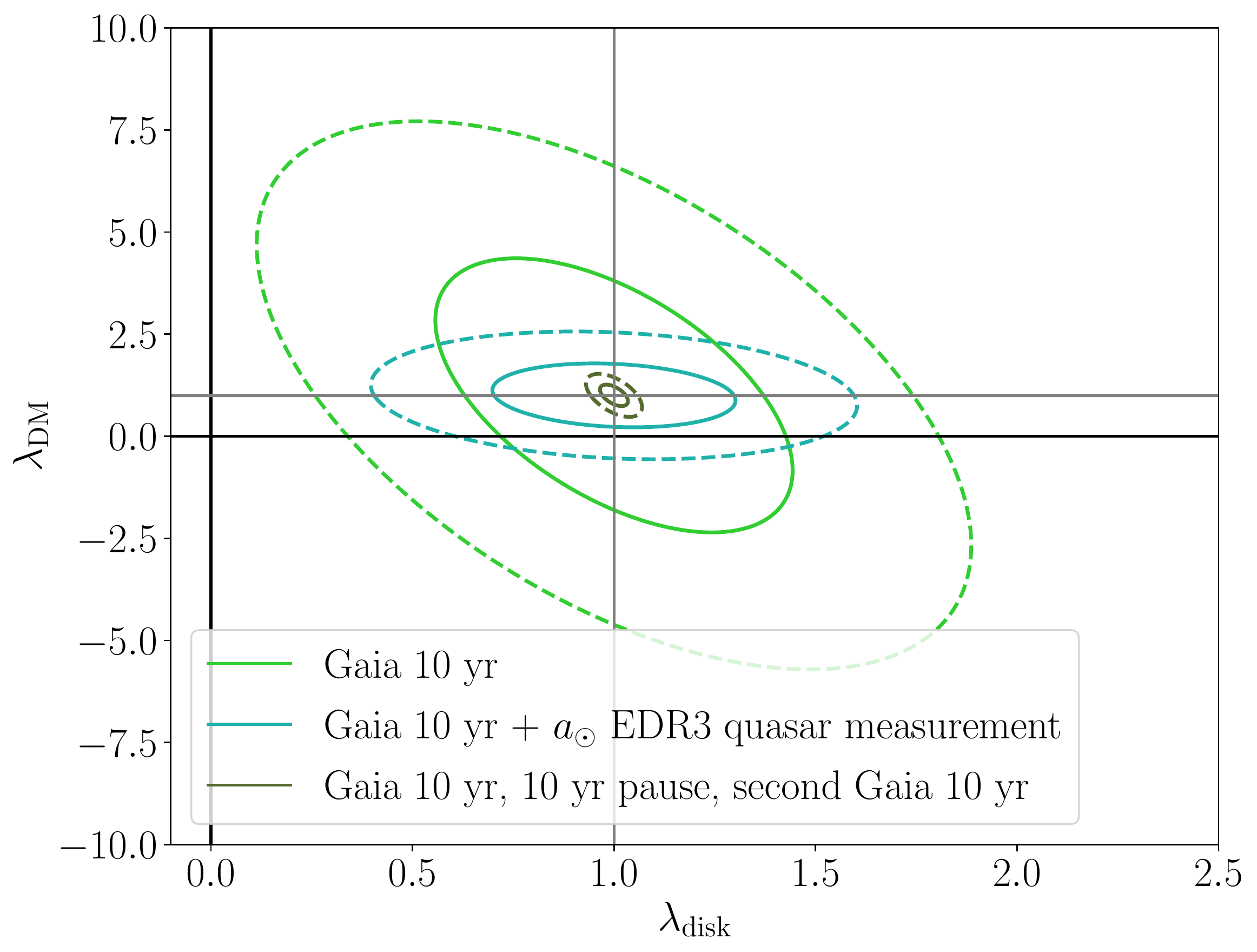}
\end{center}
\vspace{-0.4cm}
\caption{Projections for the best-fit parameter space for $\lambda_{\rm disk}$ and $\lambda_{\rm DM}$, which re-scale the disk and DM potentials, respectively, from an analysis of the Asimov data set under the signal hypothesis for 10 yrs of {\it Gaia} data. Indicated are the 1$\sigma$ (solid) and 2$\sigma$ (dashed) contours for this example, with truth values indicated by grey lines. We also show projections incorporating data from a possible survey that takes place 10 years after {\it Gaia}, as described in the text. 
\vspace{-0.4cm}} 
\label{fig:gaia_ill}
\end{figure}
We use the joint likelihood to perform inference on the model parameters ${\bm \theta}$ for the Asimov data set under the signal hypothesis. In particular, we compute the expected covariance matrix for the signal parameters by inverting the Fisher matrix evaluated for the Asimov data set. In Fig.~\ref{fig:gaia_ill} we illustrate the expected 1$\sigma$ and 2$\sigma$ contours for $\lambda_{\rm DM}$ and $\lambda_{\rm disk}$ for the 10-year {\it Gaia} data set. While the disk normalization may be measured at over 2$\sigma$ significance, the 10-year data set will not be able to infer a non-zero value of the DM normalization at more than 1$\sigma$, without incorporating additional prior information. Similarly, as shown in the SM, the bulge component is also not detectable in the 10-year data set, though future observations could be combined with {\it Gaia} to measure these components, as discussed below.

The DM contribution to the acceleration may be measured with {\it Gaia} 10-year data if prior information is incorporated into the likelihood. For example, using {\it Gaia} EDR3 the solar acceleration has been directly measured at the precision level of $\sim$7\% from the apparent motion of quasars~\cite{2020arXiv201202036G}. So far we have not incorporated this information into the likelihood; rather, the solar acceleration is self-consistently determined as part of the parameter fit, as indicated in {\it e.g.}~\eqref{eq:delta_chi2_scale}. We now consider the joint likelihood where we combine the stellar acceleration likelihood with a Gaussian likelihood for the solar acceleration, with the measured acceleration uncertainty assumed to be 7\% of the true value. Using the joint likelihood incorporating the solar acceleration measurement with the 10-year {\it Gaia} Asimov data set leads to the contours illustrated in Fig.~\ref{fig:gaia_ill}; these contours are significantly constrained relative to those without the inclusion of the prior $a_\odot$ measurement. Note that the solar acceleration measurement mostly informs the total interior matter content and thus cannot be used in isolation to disentangle the various matter components.

The above analyses indicate that {\it Gaia} may be able to directly detect the acceleration from the MW if all stars are incorporated into the joint likelihood. However, it is possible that many stars will be subject to sources of systematic uncertainty that could appreciably affect their acceleration measures. For example, source confusion is known to lead to spurious astrometric solutions for approximately 32\% of the EDR3 stars~\cite{Fabricius_2020}. The spurious solutions may be minimized by placing cuts on the parallax uncertainties. For example, requiring $\varpi / \sigma_{\varpi}  > 8.8$ $(66)$ leaves 10\% (1\%) of the EDR3 stars, but only $\sim$0.3\% ($\sim 3\cdot 10^{-7}$) of these stars are expected to be affected by source confusion~\cite{Fabricius_2020}. It is thus relevant to project sensitivity to the MW accelerations using the 10\% and 1\% data samples, as ranked by parallax uncertainties. Interestingly, we find that with the top 10\% (1\%) of the data the detection significance of the 1-parameter model, for the 10-year data set, is $q \approx 9.5$ ($q \approx 4.2$), implying that even the top 1\% data set should be able to produce over 2$\sigma$ evidence for Galactic accelerations. Additionally, zero-point offsets to the angular accelerations~\cite{2020arXiv201201742L} may present another important systematic for {\it Gaia} analyses that should be studied further. 

\mysection{Discussion}
Future surveys will drastically improve the constraining power from proper accelerations for two reasons: (i) such surveys will have improved technology, which will result in decreased measurement uncertainties, and (ii) the increase in temporal baseline relative to the {\it Gaia} data set will mean that such surveys can be combined with {\it Gaia} to better determine the accelerations. While no concrete {\it Gaia} followup survey is approved at the moment, there are many ideas for new surveys under consideration (see, {\it e.g.},~\cite{2018FrASS...5...11V} for a review), such as: the targeted {\it Theia} mission~\cite{2017arXiv170701348T}, which would be a ``point and stare" space telescope that could improve relative astrometry by $\sim$2 orders of magnitude relative to {\it Gaia}; {\it GaiaNIR}~\cite{2016arXiv160907325H}, which would be a {\it Gaia}-like mission but in the near-infrared with comparable coverage and astrometric precision; or simply repeating the {\it Gaia} mission some number of decades in the future~\cite{2018FrASS...5...11V}. However, confounding backgrounds, such as exoplanets and sub-threshold binary stars~\cite{Ravi:2018vqd,Chakrabarti:2020kco}, may become more important as statistical uncertainties shrink with future surveys, which deserves further study. 

As a concrete illustration of a future scenario, let us imagine that 10 yrs after the end of the {\it Gaia} mission, which is assumed to operate for 10 yrs, there is a second identical {\it Gaia} mission which also takes data for 10 yrs, such that the {\it Gaia} followup mission would end around 2043. The results from an analysis of the Asimov data set for such a joint data set are illustrated in Fig.~\ref{fig:gaia_ill}; the normalization of the DM halo could be constrained at over 3$\sigma$, independent of the disk and bulge normalization parameters and without the inclusion of any prior information. The evidence for accelerations in the one-parameter model would be measured at $q \approx 1.5 \times 10^3$, or approximately 40$\sigma$ significance. At 68\% containment the NFW scale radius, which was fixed at $r_s = 16$ kpc in our {\it Gaia} 10-year projection, could be treated as a model parameter and constrained to the range $r_s \in (11,26)$ kpc, at 68\% confidence, without the inclusion of prior data on $a_\odot$, and to the level $r_s \in (14.8,17.4)$ kpc including the $a_\odot$ measurement from~\cite{2020arXiv201202036G}. At this point stellar accelerations would likely become the most sensitive and robust probe of the Galactic potential and could be used to search for non-trivial DM substructure (see also~\cite{VanTilburg:2018ykj,Mishra-Sharma:2020ynk,Mondino:2020rkn}) predicted in some particle DM scenarios, such as the dark disk~\cite{Fan:2013tia,Fan:2013yva,Schutz2018}.

\mysection{Acknowledgements}
{\it We thank Nicholas Rodd for collaboration in the early stages of this work, and we also thank Vasily Belkurov, Jo Bovy, Siddharth Mishra-Sharma, and Ken Van Tilburg for useful discussions. M.B. is supported by the DOE under Award Number DESC0007968. B.R.S is supported in part by the DOE Early Career Grant DESC0019225 and by computational resources at the Lawrencium computational cluster provided by the IT Division at the Lawrence Berkeley National Laboratory, supported by the Director, Office of Science, and Office of Basic Energy Sciences, of the U.S. Department of Energy under Contract No. DE-AC02-05CH11231. K.S. was supported by a Pappalardo Fellowship in the MIT Department of Physics and by NASA through the NASA Hubble Fellowship grant HST-HF2-51470.001-A awarded by the Space Telescope Science Institute, which is operated by the Association of Universities for Research in Astronomy, Incorporated, under NASA contract NAS5-26555.
}

\bibliography{references}

\begin{thebibliography}{51}%
\makeatletter
\providecommand \@ifxundefined [1]{%
 \@ifx{#1\undefined}
}%
\providecommand \@ifnum [1]{%
 \ifnum #1\expandafter \@firstoftwo
 \else \expandafter \@secondoftwo
 \fi
}%
\providecommand \@ifx [1]{%
 \ifx #1\expandafter \@firstoftwo
 \else \expandafter \@secondoftwo
 \fi
}%
\providecommand \natexlab [1]{#1}%
\providecommand \enquote  [1]{``#1''}%
\providecommand \bibnamefont  [1]{#1}%
\providecommand \bibfnamefont [1]{#1}%
\providecommand \citenamefont [1]{#1}%
\providecommand \href@noop [0]{\@secondoftwo}%
\providecommand \href [0]{\begingroup \@sanitize@url \@href}%
\providecommand \@href[1]{\@@startlink{#1}\@@href}%
\providecommand \@@href[1]{\endgroup#1\@@endlink}%
\providecommand \@sanitize@url [0]{\catcode `\\12\catcode `\$12\catcode
  `\&12\catcode `\#12\catcode `\^12\catcode `\_12\catcode `\%12\relax}%
\providecommand \@@startlink[1]{}%
\providecommand \@@endlink[0]{}%
\providecommand \url  [0]{\begingroup\@sanitize@url \@url }%
\providecommand \@url [1]{\endgroup\@href {#1}{\urlprefix }}%
\providecommand \urlprefix  [0]{URL }%
\providecommand \Eprint [0]{\href }%
\providecommand \doibase [0]{http://dx.doi.org/}%
\providecommand \selectlanguage [0]{\@gobble}%
\providecommand \bibinfo  [0]{\@secondoftwo}%
\providecommand \bibfield  [0]{\@secondoftwo}%
\providecommand \translation [1]{[#1]}%
\providecommand \BibitemOpen [0]{}%
\providecommand \bibitemStop [0]{}%
\providecommand \bibitemNoStop [0]{.\EOS\space}%
\providecommand \EOS [0]{\spacefactor3000\relax}%
\providecommand \BibitemShut  [1]{\csname bibitem#1\endcsname}%
\let\auto@bib@innerbib\@empty
\bibitem [{\citenamefont {McKee}\ \emph {et~al.}(2015)\citenamefont {McKee},
  \citenamefont {Parravano},\ and\ \citenamefont
  {Hollenbach}}]{mckee2015stars}%
  \BibitemOpen
  \bibfield  {author} {\bibinfo {author} {\bibfnamefont {C.~F.}\ \bibnamefont
  {McKee}}, \bibinfo {author} {\bibfnamefont {A.}~\bibnamefont {Parravano}}, \
  and\ \bibinfo {author} {\bibfnamefont {D.~J.}\ \bibnamefont {Hollenbach}},\
  }\href {\doibase 10.1088/0004-637x/814/1/13} {\bibfield  {journal} {\bibinfo
  {journal} {The Astrophysical Journal}\ }\textbf {\bibinfo {volume} {814}},\
  \bibinfo {pages} {13} (\bibinfo {year} {2015})}\BibitemShut {NoStop}%
\bibitem [{\citenamefont {{Bovy}}(2017)}]{2017MNRAS.470.1360B}%
  \BibitemOpen
  \bibfield  {author} {\bibinfo {author} {\bibfnamefont {J.}~\bibnamefont
  {{Bovy}}},\ }\href {\doibase 10.1093/mnras/stx1277} {\bibfield  {journal}
  {\bibinfo  {journal} {Mon. Not. Roy. Astron. Soc.}\ }\textbf {\bibinfo
  {volume} {470}},\ \bibinfo {pages} {1360} (\bibinfo {year} {2017})},\ \Eprint
  {http://arxiv.org/abs/1704.05063} {arXiv:1704.05063 [astro-ph.GA]}
  \BibitemShut {NoStop}%
\bibitem [{\citenamefont {{Eilers}}\ \emph {et~al.}(2019)\citenamefont
  {{Eilers}}, \citenamefont {{Hogg}}, \citenamefont {{Rix}},\ and\
  \citenamefont {{Ness}}}]{2019ApJ...871..120E}%
  \BibitemOpen
  \bibfield  {author} {\bibinfo {author} {\bibfnamefont {A.-C.}\ \bibnamefont
  {{Eilers}}}, \bibinfo {author} {\bibfnamefont {D.~W.}\ \bibnamefont
  {{Hogg}}}, \bibinfo {author} {\bibfnamefont {H.-W.}\ \bibnamefont {{Rix}}}, \
  and\ \bibinfo {author} {\bibfnamefont {M.~K.}\ \bibnamefont {{Ness}}},\
  }\href {\doibase 10.3847/1538-4357/aaf648} {\bibfield  {journal} {\bibinfo
  {journal} {\apj}\ }\textbf {\bibinfo {volume} {871}},\ \bibinfo {eid} {120}
  (\bibinfo {year} {2019})},\ \Eprint {http://arxiv.org/abs/1810.09466}
  {arXiv:1810.09466 [astro-ph.GA]} \BibitemShut {NoStop}%
\bibitem [{\citenamefont {Benito}\ \emph {et~al.}(2020)\citenamefont {Benito},
  \citenamefont {Iocco},\ and\ \citenamefont {Cuoco}}]{Benito:2020lgu}%
  \BibitemOpen
  \bibfield  {author} {\bibinfo {author} {\bibfnamefont {M.}~\bibnamefont
  {Benito}}, \bibinfo {author} {\bibfnamefont {F.}~\bibnamefont {Iocco}}, \
  and\ \bibinfo {author} {\bibfnamefont {A.}~\bibnamefont {Cuoco}},\
  }\href@noop {} {\  (\bibinfo {year} {2020})},\ \Eprint
  {http://arxiv.org/abs/2009.13523} {arXiv:2009.13523 [astro-ph.GA]}
  \BibitemShut {NoStop}%
\bibitem [{\citenamefont {Kuijken}\ and\ \citenamefont
  {Gilmore}(1989)}]{kuijken1989mass}%
  \BibitemOpen
  \bibfield  {author} {\bibinfo {author} {\bibfnamefont {K.}~\bibnamefont
  {Kuijken}}\ and\ \bibinfo {author} {\bibfnamefont {G.}~\bibnamefont
  {Gilmore}},\ }\href {\doibase 10.1093/mnras/239.2.605} {\bibfield  {journal}
  {\bibinfo  {journal} {Monthly Notices of the Royal Astronomical Society}\
  }\textbf {\bibinfo {volume} {239}},\ \bibinfo {pages} {605} (\bibinfo {year}
  {1989})}\BibitemShut {NoStop}%
\bibitem [{\citenamefont {Holmberg}\ and\ \citenamefont
  {Flynn}(2000)}]{holmberg2000local}%
  \BibitemOpen
  \bibfield  {author} {\bibinfo {author} {\bibfnamefont {J.}~\bibnamefont
  {Holmberg}}\ and\ \bibinfo {author} {\bibfnamefont {C.}~\bibnamefont
  {Flynn}},\ }\href {\doibase 10.1046/j.1365-8711.2000.02905.x} {\bibfield
  {journal} {\bibinfo  {journal} {Monthly Notices of the Royal Astronomical
  Society}\ }\textbf {\bibinfo {volume} {313}},\ \bibinfo {pages} {209}
  (\bibinfo {year} {2000})}\BibitemShut {NoStop}%
\bibitem [{\citenamefont {Bovy}\ and\ \citenamefont
  {Tremaine}(2012)}]{Bovy:2012tw}%
  \BibitemOpen
  \bibfield  {author} {\bibinfo {author} {\bibfnamefont {J.}~\bibnamefont
  {Bovy}}\ and\ \bibinfo {author} {\bibfnamefont {S.}~\bibnamefont
  {Tremaine}},\ }\href {\doibase 10.1088/0004-637X/756/1/89} {\bibfield
  {journal} {\bibinfo  {journal} {Astrophys. J.}\ }\textbf {\bibinfo {volume}
  {756}},\ \bibinfo {pages} {89} (\bibinfo {year} {2012})}\BibitemShut
  {NoStop}%
\bibitem [{\citenamefont {Schutz}\ \emph {et~al.}(2018)\citenamefont {Schutz},
  \citenamefont {Lin}, \citenamefont {Safdi},\ and\ \citenamefont
  {Wu}}]{Schutz2018}%
  \BibitemOpen
  \bibfield  {author} {\bibinfo {author} {\bibfnamefont {K.}~\bibnamefont
  {Schutz}}, \bibinfo {author} {\bibfnamefont {T.}~\bibnamefont {Lin}},
  \bibinfo {author} {\bibfnamefont {B.~R.}\ \bibnamefont {Safdi}}, \ and\
  \bibinfo {author} {\bibfnamefont {C.-L.}\ \bibnamefont {Wu}},\ }\href
  {\doibase 10.1103/PhysRevLett.121.081101} {\bibfield  {journal} {\bibinfo
  {journal} {Phys. Rev. Lett.}\ }\textbf {\bibinfo {volume} {121}},\ \bibinfo
  {pages} {081101} (\bibinfo {year} {2018})}\BibitemShut {NoStop}%
\bibitem [{\citenamefont {Widmark}(2019)}]{widmark2019measuring}%
  \BibitemOpen
  \bibfield  {author} {\bibinfo {author} {\bibfnamefont {A.}~\bibnamefont
  {Widmark}},\ }\href {\doibase 10.1051/0004-6361/201834718} {\bibfield
  {journal} {\bibinfo  {journal} {Astronomy \& Astrophysics}\ }\textbf
  {\bibinfo {volume} {623}},\ \bibinfo {pages} {A30} (\bibinfo {year}
  {2019})}\BibitemShut {NoStop}%
\bibitem [{\citenamefont {de~Salas}\ \emph {et~al.}(2019)\citenamefont
  {de~Salas}, \citenamefont {Malhan}, \citenamefont {Freese}, \citenamefont
  {Hattori},\ and\ \citenamefont {Valluri}}]{deSalas:2019pee}%
  \BibitemOpen
  \bibfield  {author} {\bibinfo {author} {\bibfnamefont {P.}~\bibnamefont
  {de~Salas}}, \bibinfo {author} {\bibfnamefont {K.}~\bibnamefont {Malhan}},
  \bibinfo {author} {\bibfnamefont {K.}~\bibnamefont {Freese}}, \bibinfo
  {author} {\bibfnamefont {K.}~\bibnamefont {Hattori}}, \ and\ \bibinfo
  {author} {\bibfnamefont {M.}~\bibnamefont {Valluri}},\ }\href {\doibase
  10.1088/1475-7516/2019/10/037} {\bibfield  {journal} {\bibinfo  {journal}
  {JCAP}\ }\textbf {\bibinfo {volume} {10}},\ \bibinfo {pages} {037} (\bibinfo
  {year} {2019})},\ \Eprint {http://arxiv.org/abs/1906.06133} {arXiv:1906.06133
  [astro-ph.GA]} \BibitemShut {NoStop}%
\bibitem [{\citenamefont {Bennett}\ and\ \citenamefont
  {Bovy}(2019)}]{bennett2019vertical}%
  \BibitemOpen
  \bibfield  {author} {\bibinfo {author} {\bibfnamefont {M.}~\bibnamefont
  {Bennett}}\ and\ \bibinfo {author} {\bibfnamefont {J.}~\bibnamefont {Bovy}},\
  }\href@noop {} {\bibfield  {journal} {\bibinfo  {journal} {Monthly Notices of
  the Royal Astronomical Society}\ }\textbf {\bibinfo {volume} {482}},\
  \bibinfo {pages} {1417} (\bibinfo {year} {2019})}\BibitemShut {NoStop}%
\bibitem [{\citenamefont {{Antoja}}\ \emph {et~al.}(2018)\citenamefont
  {{Antoja}}, \citenamefont {{Helmi}}, \citenamefont {{Romero-G{\'o}mez}},
  \citenamefont {{Katz}}, \citenamefont {{Babusiaux}}, \citenamefont
  {{Drimmel}}, \citenamefont {{Evans}}, \citenamefont {{Figueras}},
  \citenamefont {{Poggio}}, \citenamefont {{Reyl{\'e}}}, \citenamefont
  {{Robin}}, \citenamefont {{Seabroke}},\ and\ \citenamefont
  {{Soubiran}}}]{2018Natur.561..360A}%
  \BibitemOpen
  \bibfield  {author} {\bibinfo {author} {\bibfnamefont {T.}~\bibnamefont
  {{Antoja}}}, \bibinfo {author} {\bibfnamefont {A.}~\bibnamefont {{Helmi}}},
  \bibinfo {author} {\bibfnamefont {M.}~\bibnamefont {{Romero-G{\'o}mez}}},
  \bibinfo {author} {\bibfnamefont {D.}~\bibnamefont {{Katz}}}, \bibinfo
  {author} {\bibfnamefont {C.}~\bibnamefont {{Babusiaux}}}, \bibinfo {author}
  {\bibfnamefont {R.}~\bibnamefont {{Drimmel}}}, \bibinfo {author}
  {\bibfnamefont {D.~W.}\ \bibnamefont {{Evans}}}, \bibinfo {author}
  {\bibfnamefont {F.}~\bibnamefont {{Figueras}}}, \bibinfo {author}
  {\bibfnamefont {E.}~\bibnamefont {{Poggio}}}, \bibinfo {author}
  {\bibfnamefont {C.}~\bibnamefont {{Reyl{\'e}}}}, \bibinfo {author}
  {\bibfnamefont {A.~C.}\ \bibnamefont {{Robin}}}, \bibinfo {author}
  {\bibfnamefont {G.}~\bibnamefont {{Seabroke}}}, \ and\ \bibinfo {author}
  {\bibfnamefont {C.}~\bibnamefont {{Soubiran}}},\ }\href {\doibase
  10.1038/s41586-018-0510-7} {\bibfield  {journal} {\bibinfo  {journal} {\nat}\
  }\textbf {\bibinfo {volume} {561}},\ \bibinfo {pages} {360} (\bibinfo {year}
  {2018})},\ \Eprint {http://arxiv.org/abs/1804.10196} {arXiv:1804.10196
  [astro-ph.GA]} \BibitemShut {NoStop}%
\bibitem [{\citenamefont {Helmi}\ \emph {et~al.}(2018)\citenamefont {Helmi},
  \citenamefont {Babusiaux}, \citenamefont {Koppelman}, \citenamefont
  {Massari}, \citenamefont {Veljanoski},\ and\ \citenamefont
  {Brown}}]{helmi2018merger}%
  \BibitemOpen
  \bibfield  {author} {\bibinfo {author} {\bibfnamefont {A.}~\bibnamefont
  {Helmi}}, \bibinfo {author} {\bibfnamefont {C.}~\bibnamefont {Babusiaux}},
  \bibinfo {author} {\bibfnamefont {H.~H.}\ \bibnamefont {Koppelman}}, \bibinfo
  {author} {\bibfnamefont {D.}~\bibnamefont {Massari}}, \bibinfo {author}
  {\bibfnamefont {J.}~\bibnamefont {Veljanoski}}, \ and\ \bibinfo {author}
  {\bibfnamefont {A.~G.}\ \bibnamefont {Brown}},\ }\href@noop {} {\bibfield
  {journal} {\bibinfo  {journal} {Nature}\ }\textbf {\bibinfo {volume} {563}},\
  \bibinfo {pages} {85} (\bibinfo {year} {2018})}\BibitemShut {NoStop}%
\bibitem [{\citenamefont {Ravi}\ \emph {et~al.}(2019)\citenamefont {Ravi},
  \citenamefont {Langellier}, \citenamefont {Phillips}, \citenamefont
  {Buschmann}, \citenamefont {Safdi},\ and\ \citenamefont
  {Walsworth}}]{Ravi:2018vqd}%
  \BibitemOpen
  \bibfield  {author} {\bibinfo {author} {\bibfnamefont {A.}~\bibnamefont
  {Ravi}}, \bibinfo {author} {\bibfnamefont {N.}~\bibnamefont {Langellier}},
  \bibinfo {author} {\bibfnamefont {D.~F.}\ \bibnamefont {Phillips}}, \bibinfo
  {author} {\bibfnamefont {M.}~\bibnamefont {Buschmann}}, \bibinfo {author}
  {\bibfnamefont {B.~R.}\ \bibnamefont {Safdi}}, \ and\ \bibinfo {author}
  {\bibfnamefont {R.~L.}\ \bibnamefont {Walsworth}},\ }\href {\doibase
  10.1103/PhysRevLett.123.091101} {\bibfield  {journal} {\bibinfo  {journal}
  {Phys. Rev. Lett.}\ }\textbf {\bibinfo {volume} {123}},\ \bibinfo {pages}
  {091101} (\bibinfo {year} {2019})},\ \Eprint
  {http://arxiv.org/abs/1812.07578} {arXiv:1812.07578 [astro-ph.GA]}
  \BibitemShut {NoStop}%
\bibitem [{\citenamefont {Silverwood}\ and\ \citenamefont
  {Easther}(2019)}]{Silverwood:2018qra}%
  \BibitemOpen
  \bibfield  {author} {\bibinfo {author} {\bibfnamefont {H.}~\bibnamefont
  {Silverwood}}\ and\ \bibinfo {author} {\bibfnamefont {R.}~\bibnamefont
  {Easther}},\ }\href {\doibase 10.1017/pasa.2019.25} {\bibfield  {journal}
  {\bibinfo  {journal} {Publ. Astron. Soc. Austral.}\ }\textbf {\bibinfo
  {volume} {36}},\ \bibinfo {pages} {e038} (\bibinfo {year} {2019})},\ \Eprint
  {http://arxiv.org/abs/1812.07581} {arXiv:1812.07581 [astro-ph.GA]}
  \BibitemShut {NoStop}%
\bibitem [{\citenamefont {Chakrabarti}\ \emph
  {et~al.}(2020{\natexlab{a}})\citenamefont {Chakrabarti} \emph
  {et~al.}}]{Chakrabarti:2020kco}%
  \BibitemOpen
  \bibfield  {author} {\bibinfo {author} {\bibfnamefont {S.}~\bibnamefont
  {Chakrabarti}} \emph {et~al.},\ }\href@noop {} {\  (\bibinfo {year}
  {2020}{\natexlab{a}})},\ \Eprint {http://arxiv.org/abs/2007.15097}
  {arXiv:2007.15097 [astro-ph.GA]} \BibitemShut {NoStop}%
\bibitem [{\citenamefont {Phillips}\ \emph {et~al.}(2020)\citenamefont
  {Phillips}, \citenamefont {Ravi}, \citenamefont {Ebadi},\ and\ \citenamefont
  {Walsworth}}]{Phillips:2020xmf}%
  \BibitemOpen
  \bibfield  {author} {\bibinfo {author} {\bibfnamefont {D.~F.}\ \bibnamefont
  {Phillips}}, \bibinfo {author} {\bibfnamefont {A.}~\bibnamefont {Ravi}},
  \bibinfo {author} {\bibfnamefont {R.}~\bibnamefont {Ebadi}}, \ and\ \bibinfo
  {author} {\bibfnamefont {R.~L.}\ \bibnamefont {Walsworth}},\ }\href@noop {}
  {\  (\bibinfo {year} {2020})},\ \Eprint {http://arxiv.org/abs/2008.13052}
  {arXiv:2008.13052 [astro-ph.GA]} \BibitemShut {NoStop}%
\bibitem [{\citenamefont {Chakrabarti}\ \emph
  {et~al.}(2020{\natexlab{b}})\citenamefont {Chakrabarti}, \citenamefont
  {Chang}, \citenamefont {Lam}, \citenamefont {Vigeland},\ and\ \citenamefont
  {Quillen}}]{Chakrabarti:2020abx}%
  \BibitemOpen
  \bibfield  {author} {\bibinfo {author} {\bibfnamefont {S.}~\bibnamefont
  {Chakrabarti}}, \bibinfo {author} {\bibfnamefont {P.}~\bibnamefont {Chang}},
  \bibinfo {author} {\bibfnamefont {M.~T.}\ \bibnamefont {Lam}}, \bibinfo
  {author} {\bibfnamefont {S.~J.}\ \bibnamefont {Vigeland}}, \ and\ \bibinfo
  {author} {\bibfnamefont {A.~C.}\ \bibnamefont {Quillen}},\ }\href@noop {} {\
  (\bibinfo {year} {2020}{\natexlab{b}})},\ \Eprint
  {http://arxiv.org/abs/2010.04018} {arXiv:2010.04018 [astro-ph.GA]}
  \BibitemShut {NoStop}%
\bibitem [{\citenamefont {Bovy}(2020)}]{bovy2020purely}%
  \BibitemOpen
  \bibfield  {author} {\bibinfo {author} {\bibfnamefont {J.}~\bibnamefont
  {Bovy}},\ }\href@noop {} {\bibfield  {journal} {\bibinfo  {journal} {arXiv
  preprint}\ } (\bibinfo {year} {2020})},\ \Eprint
  {http://arxiv.org/abs/2012.02169} {arXiv:2012.02169 [astro-ph.GA]}
  \BibitemShut {NoStop}%
\bibitem [{\citenamefont {{Gaia Collaboration}}\ \emph
  {et~al.}(2020)\citenamefont {{Gaia Collaboration}}, \citenamefont
  {{Klioner}}, \citenamefont {{Mignard}}, \citenamefont {{Lindegren}},
  \citenamefont {{Bastian}}, \citenamefont {{McMillan}}, \citenamefont
  {{Hern{\'a}ndez}}, \citenamefont {{Hobbs}}, \citenamefont {{Ramos-Lerate}},
  \citenamefont {{Biermann}}, \citenamefont {{Bombrun}}, \citenamefont {{de
  Torres}}, \citenamefont {{Gerlach}}, \citenamefont {{Geyer}}, \citenamefont
  {{Hilger}}, \citenamefont {{Lammers}} \emph {et~al.}}]{2020arXiv201202036G}%
  \BibitemOpen
  \bibfield  {author} {\bibinfo {author} {\bibnamefont {{Gaia Collaboration}}},
  \bibinfo {author} {\bibfnamefont {S.~A.}\ \bibnamefont {{Klioner}}}, \bibinfo
  {author} {\bibfnamefont {F.}~\bibnamefont {{Mignard}}}, \bibinfo {author}
  {\bibfnamefont {L.}~\bibnamefont {{Lindegren}}}, \bibinfo {author}
  {\bibfnamefont {U.}~\bibnamefont {{Bastian}}}, \bibinfo {author}
  {\bibfnamefont {P.~J.}\ \bibnamefont {{McMillan}}}, \bibinfo {author}
  {\bibfnamefont {J.}~\bibnamefont {{Hern{\'a}ndez}}}, \bibinfo {author}
  {\bibfnamefont {D.}~\bibnamefont {{Hobbs}}}, \bibinfo {author} {\bibfnamefont
  {M.}~\bibnamefont {{Ramos-Lerate}}}, \bibinfo {author} {\bibfnamefont
  {M.}~\bibnamefont {{Biermann}}}, \bibinfo {author} {\bibfnamefont
  {A.}~\bibnamefont {{Bombrun}}}, \bibinfo {author} {\bibfnamefont
  {A.}~\bibnamefont {{de Torres}}}, \bibinfo {author} {\bibfnamefont
  {E.}~\bibnamefont {{Gerlach}}}, \bibinfo {author} {\bibfnamefont
  {R.}~\bibnamefont {{Geyer}}}, \bibinfo {author} {\bibfnamefont
  {T.}~\bibnamefont {{Hilger}}}, \bibinfo {author} {\bibnamefont {{Lammers}}},
  \emph {et~al.},\ }\href@noop {} {\bibfield  {journal} {\bibinfo  {journal}
  {arXiv e-prints}\ } (\bibinfo {year} {2020})},\ \Eprint
  {http://arxiv.org/abs/2012.02036} {arXiv:2012.02036 [astro-ph.GA]}
  \BibitemShut {NoStop}%
\bibitem [{\citenamefont {Brown}\ \emph {et~al.}(2018)\citenamefont {Brown},
  \citenamefont {Vallenari}, \citenamefont {Prusti}, \citenamefont
  {de~Bruijne}, \citenamefont {Babusiaux}, \citenamefont {Bailer-Jones},
  \citenamefont {Biermann}, \citenamefont {Evans}, \citenamefont {Eyer},
  \citenamefont {Jansen} \emph {et~al.}}]{Brown2018}%
  \BibitemOpen
  \bibfield  {author} {\bibinfo {author} {\bibfnamefont {A.~G.~A.}\
  \bibnamefont {Brown}}, \bibinfo {author} {\bibfnamefont {A.}~\bibnamefont
  {Vallenari}}, \bibinfo {author} {\bibfnamefont {T.}~\bibnamefont {Prusti}},
  \bibinfo {author} {\bibfnamefont {J.~H.~J.}\ \bibnamefont {de~Bruijne}},
  \bibinfo {author} {\bibfnamefont {C.}~\bibnamefont {Babusiaux}}, \bibinfo
  {author} {\bibfnamefont {C.~A.~L.}\ \bibnamefont {Bailer-Jones}}, \bibinfo
  {author} {\bibfnamefont {M.}~\bibnamefont {Biermann}}, \bibinfo {author}
  {\bibfnamefont {D.~W.}\ \bibnamefont {Evans}}, \bibinfo {author}
  {\bibfnamefont {L.}~\bibnamefont {Eyer}}, \bibinfo {author} {\bibfnamefont
  {F.}~\bibnamefont {Jansen}},  \emph {et~al.},\ }\href {\doibase
  10.1051/0004-6361/201833051} {\bibfield  {journal} {\bibinfo  {journal}
  {Astron. Astrophys.}\ }\textbf {\bibinfo {volume} {616}},\ \bibinfo {pages}
  {A1} (\bibinfo {year} {2018})}\BibitemShut {NoStop}%
\bibitem [{\citenamefont {{Lindegren}}\ \emph {et~al.}(2018)\citenamefont
  {{Lindegren}}, \citenamefont {{Hern{\'a}ndez}}, \citenamefont {{Bombrun}},
  \citenamefont {{Klioner}}, \citenamefont {{Bastian}}, \citenamefont
  {{Ramos-Lerate}}, \citenamefont {{de Torres}}, \citenamefont
  {{Steidelm{\"u}ller}}, \citenamefont {{Stephenson}}, \citenamefont {{Hobbs}}
  \emph {et~al.}}]{2018A&A...616A...2L}%
  \BibitemOpen
  \bibfield  {author} {\bibinfo {author} {\bibfnamefont {L.}~\bibnamefont
  {{Lindegren}}}, \bibinfo {author} {\bibfnamefont {J.}~\bibnamefont
  {{Hern{\'a}ndez}}}, \bibinfo {author} {\bibfnamefont {A.}~\bibnamefont
  {{Bombrun}}}, \bibinfo {author} {\bibfnamefont {S.}~\bibnamefont
  {{Klioner}}}, \bibinfo {author} {\bibfnamefont {U.}~\bibnamefont
  {{Bastian}}}, \bibinfo {author} {\bibfnamefont {M.}~\bibnamefont
  {{Ramos-Lerate}}}, \bibinfo {author} {\bibfnamefont {A.}~\bibnamefont {{de
  Torres}}}, \bibinfo {author} {\bibfnamefont {H.}~\bibnamefont
  {{Steidelm{\"u}ller}}}, \bibinfo {author} {\bibfnamefont {C.}~\bibnamefont
  {{Stephenson}}}, \bibinfo {author} {\bibfnamefont {D.}~\bibnamefont
  {{Hobbs}}},  \emph {et~al.},\ }\href {\doibase 10.1051/0004-6361/201832727}
  {\bibfield  {journal} {\bibinfo  {journal} {Astron. Astrophys.}\ }\textbf
  {\bibinfo {volume} {616}},\ \bibinfo {eid} {A2} (\bibinfo {year} {2018})},\
  \Eprint {http://arxiv.org/abs/1804.09366} {arXiv:1804.09366 [astro-ph.IM]}
  \BibitemShut {NoStop}%
\bibitem [{199(1997)}]{1997ESASP1200.....E}%
  \BibitemOpen
  \href@noop {} {\emph {\bibinfo {title} {ESA Special Publication}}},\ \bibinfo
  {series} {ESA Special Publication}, Vol.\ \bibinfo {volume} {1200}\ (\bibinfo
  {year} {1997})\BibitemShut {NoStop}%
\bibitem [{\citenamefont {{Brandt}}(2018)}]{2018ApJS..239...31B}%
  \BibitemOpen
  \bibfield  {author} {\bibinfo {author} {\bibfnamefont {T.~D.}\ \bibnamefont
  {{Brandt}}},\ }\href {\doibase 10.3847/1538-4365/aaec06} {\bibfield
  {journal} {\bibinfo  {journal} {ApJS}\ }\textbf {\bibinfo {volume} {239}},\
  \bibinfo {eid} {31} (\bibinfo {year} {2018})},\ \Eprint
  {http://arxiv.org/abs/1811.07283} {arXiv:1811.07283 [astro-ph.SR]}
  \BibitemShut {NoStop}%
\bibitem [{\citenamefont {Van~Tilburg}\ \emph {et~al.}(2018)\citenamefont
  {Van~Tilburg}, \citenamefont {Taki},\ and\ \citenamefont
  {Weiner}}]{VanTilburg:2018ykj}%
  \BibitemOpen
  \bibfield  {author} {\bibinfo {author} {\bibfnamefont {K.}~\bibnamefont
  {Van~Tilburg}}, \bibinfo {author} {\bibfnamefont {A.-M.}\ \bibnamefont
  {Taki}}, \ and\ \bibinfo {author} {\bibfnamefont {N.}~\bibnamefont
  {Weiner}},\ }\href {\doibase 10.1088/1475-7516/2018/07/041} {\bibfield
  {journal} {\bibinfo  {journal} {J. Cosmol. Astropart. Phys.}\ }\textbf
  {\bibinfo {volume} {1807}},\ \bibinfo {pages} {041} (\bibinfo {year}
  {2018})}\BibitemShut {NoStop}%
\bibitem [{\citenamefont {Abuter}\ \emph {et~al.}(2018)\citenamefont {Abuter},
  \citenamefont {Amorim}, \citenamefont {Anugu}, \citenamefont {Baub{\"o}ck},
  \citenamefont {Benisty}, \citenamefont {Berger}, \citenamefont {Blind},
  \citenamefont {Bonnet}, \citenamefont {Brandner},\ and\ \citenamefont
  {et~al.}}]{du2018detection}%
  \BibitemOpen
  \bibfield  {author} {\bibinfo {author} {\bibfnamefont {R.}~\bibnamefont
  {Abuter}}, \bibinfo {author} {\bibfnamefont {A.}~\bibnamefont {Amorim}},
  \bibinfo {author} {\bibfnamefont {N.}~\bibnamefont {Anugu}}, \bibinfo
  {author} {\bibfnamefont {M.}~\bibnamefont {Baub{\"o}ck}}, \bibinfo {author}
  {\bibfnamefont {M.}~\bibnamefont {Benisty}}, \bibinfo {author} {\bibfnamefont
  {J.~P.}\ \bibnamefont {Berger}}, \bibinfo {author} {\bibfnamefont
  {N.}~\bibnamefont {Blind}}, \bibinfo {author} {\bibfnamefont
  {H.}~\bibnamefont {Bonnet}}, \bibinfo {author} {\bibfnamefont
  {W.}~\bibnamefont {Brandner}}, \ and\ \bibinfo {author} {\bibnamefont
  {et~al.}},\ }\href {\doibase 10.1051/0004-6361/201833718} {\bibfield
  {journal} {\bibinfo  {journal} {Astronomy \& Astrophysics}\ }\textbf
  {\bibinfo {volume} {615}},\ \bibinfo {pages} {L15} (\bibinfo {year}
  {2018})}\BibitemShut {NoStop}%
\bibitem [{\citenamefont {Abuter}\ \emph {et~al.}(2020)\citenamefont {Abuter},
  \citenamefont {Amorim}, \citenamefont {Baub{\"o}ck}, \citenamefont {Berger},
  \citenamefont {Bonnet}, \citenamefont {Brandner}, \citenamefont {Cardoso},
  \citenamefont {Cl{\'e}net}, \citenamefont {de~Zeeuw},\ and\ \citenamefont
  {et~al.}}]{abuter2020arxiv}%
  \BibitemOpen
  \bibfield  {author} {\bibinfo {author} {\bibfnamefont {R.}~\bibnamefont
  {Abuter}}, \bibinfo {author} {\bibfnamefont {A.}~\bibnamefont {Amorim}},
  \bibinfo {author} {\bibfnamefont {M.}~\bibnamefont {Baub{\"o}ck}}, \bibinfo
  {author} {\bibfnamefont {J.~P.}\ \bibnamefont {Berger}}, \bibinfo {author}
  {\bibfnamefont {H.}~\bibnamefont {Bonnet}}, \bibinfo {author} {\bibfnamefont
  {W.}~\bibnamefont {Brandner}}, \bibinfo {author} {\bibfnamefont
  {V.}~\bibnamefont {Cardoso}}, \bibinfo {author} {\bibfnamefont
  {Y.}~\bibnamefont {Cl{\'e}net}}, \bibinfo {author} {\bibfnamefont {P.~T.}\
  \bibnamefont {de~Zeeuw}}, \ and\ \bibinfo {author} {\bibnamefont {et~al.}},\
  }\href {\doibase 10.1051/0004-6361/202037813} {\bibfield  {journal} {\bibinfo
   {journal} {Astronomy \& Astrophysics}\ }\textbf {\bibinfo {volume} {636}},\
  \bibinfo {pages} {L5} (\bibinfo {year} {2020})}\BibitemShut {NoStop}%
\bibitem [{\citenamefont {{Gravity Collaboration}}\ \emph
  {et~al.}(2019)\citenamefont {{Gravity Collaboration}}, \citenamefont
  {{Abuter}}, \citenamefont {{Amorim}}, \citenamefont {{Baub{\"o}ck}},
  \citenamefont {{Berger}}, \citenamefont {{Bonnet}}, \citenamefont
  {{Brandner}}, \citenamefont {{Cl{\'e}net}}, \citenamefont {{Coud{\'e} Du
  Foresto}}, \citenamefont {{de Zeeuw}}, \citenamefont {{Dexter}},
  \citenamefont {{Duvert}}, \citenamefont {{Eckart}}, \citenamefont
  {{Eisenhauer}}, \citenamefont {{F{\"o}rster Schreiber}} \emph
  {et~al.}}]{2019A&A...625L..10G}%
  \BibitemOpen
  \bibfield  {author} {\bibinfo {author} {\bibnamefont {{Gravity
  Collaboration}}}, \bibinfo {author} {\bibfnamefont {R.}~\bibnamefont
  {{Abuter}}}, \bibinfo {author} {\bibfnamefont {A.}~\bibnamefont {{Amorim}}},
  \bibinfo {author} {\bibfnamefont {M.}~\bibnamefont {{Baub{\"o}ck}}}, \bibinfo
  {author} {\bibfnamefont {J.~P.}\ \bibnamefont {{Berger}}}, \bibinfo {author}
  {\bibfnamefont {H.}~\bibnamefont {{Bonnet}}}, \bibinfo {author}
  {\bibfnamefont {W.}~\bibnamefont {{Brandner}}}, \bibinfo {author}
  {\bibfnamefont {Y.}~\bibnamefont {{Cl{\'e}net}}}, \bibinfo {author}
  {\bibfnamefont {V.}~\bibnamefont {{Coud{\'e} Du Foresto}}}, \bibinfo {author}
  {\bibfnamefont {P.~T.}\ \bibnamefont {{de Zeeuw}}}, \bibinfo {author}
  {\bibfnamefont {J.}~\bibnamefont {{Dexter}}}, \bibinfo {author}
  {\bibfnamefont {G.}~\bibnamefont {{Duvert}}}, \bibinfo {author}
  {\bibfnamefont {A.}~\bibnamefont {{Eckart}}}, \bibinfo {author}
  {\bibfnamefont {F.}~\bibnamefont {{Eisenhauer}}}, \bibinfo {author}
  {\bibfnamefont {N.~M.}\ \bibnamefont {{F{\"o}rster Schreiber}}},  \emph
  {et~al.},\ }\href {\doibase 10.1051/0004-6361/201935656} {\bibfield
  {journal} {\bibinfo  {journal} {\aap}\ }\textbf {\bibinfo {volume} {625}},\
  \bibinfo {eid} {L10} (\bibinfo {year} {2019})},\ \Eprint
  {http://arxiv.org/abs/1904.05721} {arXiv:1904.05721 [astro-ph.GA]}
  \BibitemShut {NoStop}%
\bibitem [{\citenamefont {{Erkal}}\ \emph {et~al.}(2019)\citenamefont
  {{Erkal}}, \citenamefont {{Belokurov}}, \citenamefont {{Laporte}},
  \citenamefont {{Koposov}}, \citenamefont {{Li}}, \citenamefont {{Grillmair}},
  \citenamefont {{Kallivayalil}}, \citenamefont {{Price-Whelan}}, \citenamefont
  {{Evans}}, \citenamefont {{Hawkins}}, \citenamefont {{Hendel}}, \citenamefont
  {{Mateu}}, \citenamefont {{Navarro}}, \citenamefont {{del Pino}},
  \citenamefont {{Slater}}, \citenamefont {{Sohn}},\ and\ \citenamefont
  {{Orphan Aspen Treasury Collaboration}}}]{2019MNRAS.487.2685E}%
  \BibitemOpen
  \bibfield  {author} {\bibinfo {author} {\bibfnamefont {D.}~\bibnamefont
  {{Erkal}}}, \bibinfo {author} {\bibfnamefont {V.}~\bibnamefont
  {{Belokurov}}}, \bibinfo {author} {\bibfnamefont {C.~F.~P.}\ \bibnamefont
  {{Laporte}}}, \bibinfo {author} {\bibfnamefont {S.~E.}\ \bibnamefont
  {{Koposov}}}, \bibinfo {author} {\bibfnamefont {T.~S.}\ \bibnamefont {{Li}}},
  \bibinfo {author} {\bibfnamefont {C.~J.}\ \bibnamefont {{Grillmair}}},
  \bibinfo {author} {\bibfnamefont {N.}~\bibnamefont {{Kallivayalil}}},
  \bibinfo {author} {\bibfnamefont {A.~M.}\ \bibnamefont {{Price-Whelan}}},
  \bibinfo {author} {\bibfnamefont {N.~W.}\ \bibnamefont {{Evans}}}, \bibinfo
  {author} {\bibfnamefont {K.}~\bibnamefont {{Hawkins}}}, \bibinfo {author}
  {\bibfnamefont {D.}~\bibnamefont {{Hendel}}}, \bibinfo {author}
  {\bibfnamefont {C.}~\bibnamefont {{Mateu}}}, \bibinfo {author} {\bibfnamefont
  {J.~F.}\ \bibnamefont {{Navarro}}}, \bibinfo {author} {\bibfnamefont
  {A.}~\bibnamefont {{del Pino}}}, \bibinfo {author} {\bibfnamefont {C.~T.}\
  \bibnamefont {{Slater}}}, \bibinfo {author} {\bibfnamefont {S.~T.}\
  \bibnamefont {{Sohn}}}, \ and\ \bibinfo {author} {\bibnamefont {{Orphan Aspen
  Treasury Collaboration}}},\ }\href {\doibase 10.1093/mnras/stz1371}
  {\bibfield  {journal} {\bibinfo  {journal} {Mon. Not. Roy. Astron. Soc.}\
  }\textbf {\bibinfo {volume} {487}},\ \bibinfo {pages} {2685} (\bibinfo {year}
  {2019})},\ \Eprint {http://arxiv.org/abs/1812.08192} {arXiv:1812.08192
  [astro-ph.GA]} \BibitemShut {NoStop}%
\bibitem [{\citenamefont {{Erkal}}\ \emph {et~al.}(2020)\citenamefont
  {{Erkal}}, \citenamefont {{Deason}}, \citenamefont {{Belokurov}},
  \citenamefont {{Xue}}, \citenamefont {{Koposov}}, \citenamefont {{Bird}},
  \citenamefont {{Liu}}, \citenamefont {{Simion}}, \citenamefont {{Yang}},
  \citenamefont {{Zhang}},\ and\ \citenamefont {{Zhao}}}]{2020arXiv201013789E}%
  \BibitemOpen
  \bibfield  {author} {\bibinfo {author} {\bibfnamefont {D.}~\bibnamefont
  {{Erkal}}}, \bibinfo {author} {\bibfnamefont {A.~J.}\ \bibnamefont
  {{Deason}}}, \bibinfo {author} {\bibfnamefont {V.}~\bibnamefont
  {{Belokurov}}}, \bibinfo {author} {\bibfnamefont {X.-X.}\ \bibnamefont
  {{Xue}}}, \bibinfo {author} {\bibfnamefont {S.~E.}\ \bibnamefont
  {{Koposov}}}, \bibinfo {author} {\bibfnamefont {S.~A.}\ \bibnamefont
  {{Bird}}}, \bibinfo {author} {\bibfnamefont {C.}~\bibnamefont {{Liu}}},
  \bibinfo {author} {\bibfnamefont {I.~T.}\ \bibnamefont {{Simion}}}, \bibinfo
  {author} {\bibfnamefont {C.}~\bibnamefont {{Yang}}}, \bibinfo {author}
  {\bibfnamefont {L.}~\bibnamefont {{Zhang}}}, \ and\ \bibinfo {author}
  {\bibfnamefont {G.}~\bibnamefont {{Zhao}}},\ }\href@noop {} {\bibfield
  {journal} {\bibinfo  {journal} {arXiv e-prints}\ ,\ \bibinfo {eid}
  {arXiv:2010.13789}} (\bibinfo {year} {2020})},\ \Eprint
  {http://arxiv.org/abs/2010.13789} {arXiv:2010.13789 [astro-ph.GA]}
  \BibitemShut {NoStop}%
\bibitem [{\citenamefont {{Vasiliev}}\ \emph {et~al.}(2021)\citenamefont
  {{Vasiliev}}, \citenamefont {{Belokurov}},\ and\ \citenamefont
  {{Erkal}}}]{2021MNRAS.501.2279V}%
  \BibitemOpen
  \bibfield  {author} {\bibinfo {author} {\bibfnamefont {E.}~\bibnamefont
  {{Vasiliev}}}, \bibinfo {author} {\bibfnamefont {V.}~\bibnamefont
  {{Belokurov}}}, \ and\ \bibinfo {author} {\bibfnamefont {D.}~\bibnamefont
  {{Erkal}}},\ }\href {\doibase 10.1093/mnras/staa3673} {\bibfield  {journal}
  {\bibinfo  {journal} {Mon. Not. Roy. Astron. Soc.}\ }\textbf {\bibinfo
  {volume} {501}},\ \bibinfo {pages} {2279} (\bibinfo {year} {2021})},\ \Eprint
  {http://arxiv.org/abs/2009.10726} {arXiv:2009.10726 [astro-ph.GA]}
  \BibitemShut {NoStop}%
\bibitem [{\citenamefont {{Hernquist}}(1990)}]{1990ApJ...356..359H}%
  \BibitemOpen
  \bibfield  {author} {\bibinfo {author} {\bibfnamefont {L.}~\bibnamefont
  {{Hernquist}}},\ }\href {\doibase 10.1086/168845} {\bibfield  {journal}
  {\bibinfo  {journal} {ApJ}\ }\textbf {\bibinfo {volume} {356}},\ \bibinfo
  {pages} {359} (\bibinfo {year} {1990})}\BibitemShut {NoStop}%
\bibitem [{\citenamefont {{Bovy}}\ and\ \citenamefont
  {{Rix}}(2013)}]{2013ApJ...779..115B}%
  \BibitemOpen
  \bibfield  {author} {\bibinfo {author} {\bibfnamefont {J.}~\bibnamefont
  {{Bovy}}}\ and\ \bibinfo {author} {\bibfnamefont {H.-W.}\ \bibnamefont
  {{Rix}}},\ }\href {\doibase 10.1088/0004-637X/779/2/115} {\bibfield
  {journal} {\bibinfo  {journal} {ApJ}\ }\textbf {\bibinfo {volume} {779}},\
  \bibinfo {eid} {115} (\bibinfo {year} {2013})},\ \Eprint
  {http://arxiv.org/abs/1309.0809} {arXiv:1309.0809 [astro-ph.GA]} \BibitemShut
  {NoStop}%
\bibitem [{\citenamefont {{Bovy}}(2015)}]{2015ApJS..216...29B}%
  \BibitemOpen
  \bibfield  {author} {\bibinfo {author} {\bibfnamefont {J.}~\bibnamefont
  {{Bovy}}},\ }\href {\doibase 10.1088/0067-0049/216/2/29} {\bibfield
  {journal} {\bibinfo  {journal} {ApJS}\ }\textbf {\bibinfo {volume} {216}},\
  \bibinfo {eid} {29} (\bibinfo {year} {2015})},\ \Eprint
  {http://arxiv.org/abs/1412.3451} {arXiv:1412.3451 [astro-ph.GA]} \BibitemShut
  {NoStop}%
\bibitem [{\citenamefont {{Miyamoto}}\ and\ \citenamefont
  {{Nagai}}(1975)}]{1975PASJ...27..533M}%
  \BibitemOpen
  \bibfield  {author} {\bibinfo {author} {\bibfnamefont {M.}~\bibnamefont
  {{Miyamoto}}}\ and\ \bibinfo {author} {\bibfnamefont {R.}~\bibnamefont
  {{Nagai}}},\ }\href@noop {} {\bibfield  {journal} {\bibinfo  {journal}
  {PASJ}\ }\textbf {\bibinfo {volume} {27}},\ \bibinfo {pages} {533} (\bibinfo
  {year} {1975})}\BibitemShut {NoStop}%
\bibitem [{\citenamefont {Cautun}\ \emph {et~al.}(2020)\citenamefont {Cautun},
  \citenamefont {Ben{\'\i}tez-Llambay}, \citenamefont {Deason}, \citenamefont
  {Frenk}, \citenamefont {Fattahi}, \citenamefont {G{\'o}mez}, \citenamefont
  {Grand}, \citenamefont {Oman}, \citenamefont {Navarro},\ and\ \citenamefont
  {Simpson}}]{cautun2020milky}%
  \BibitemOpen
  \bibfield  {author} {\bibinfo {author} {\bibfnamefont {M.}~\bibnamefont
  {Cautun}}, \bibinfo {author} {\bibfnamefont {A.}~\bibnamefont
  {Ben{\'\i}tez-Llambay}}, \bibinfo {author} {\bibfnamefont {A.~J.}\
  \bibnamefont {Deason}}, \bibinfo {author} {\bibfnamefont {C.~S.}\
  \bibnamefont {Frenk}}, \bibinfo {author} {\bibfnamefont {A.}~\bibnamefont
  {Fattahi}}, \bibinfo {author} {\bibfnamefont {F.~A.}\ \bibnamefont
  {G{\'o}mez}}, \bibinfo {author} {\bibfnamefont {R.~J.~J.}\ \bibnamefont
  {Grand}}, \bibinfo {author} {\bibfnamefont {K.~A.}\ \bibnamefont {Oman}},
  \bibinfo {author} {\bibfnamefont {J.~F.}\ \bibnamefont {Navarro}}, \ and\
  \bibinfo {author} {\bibfnamefont {C.~M.}\ \bibnamefont {Simpson}},\ }\href
  {\doibase 10.1093/mnras/staa1017} {\bibfield  {journal} {\bibinfo  {journal}
  {Monthly Notices of the Royal Astronomical Society}\ }\textbf {\bibinfo
  {volume} {494}},\ \bibinfo {pages} {4291} (\bibinfo {year}
  {2020})}\BibitemShut {NoStop}%
\bibitem [{\citenamefont {Nitschai}\ \emph {et~al.}(2020)\citenamefont
  {Nitschai}, \citenamefont {Cappellari},\ and\ \citenamefont
  {Neumayer}}]{nitschai2020first}%
  \BibitemOpen
  \bibfield  {author} {\bibinfo {author} {\bibfnamefont {M.~S.}\ \bibnamefont
  {Nitschai}}, \bibinfo {author} {\bibfnamefont {M.}~\bibnamefont
  {Cappellari}}, \ and\ \bibinfo {author} {\bibfnamefont {N.}~\bibnamefont
  {Neumayer}},\ }\href {\doibase 10.1093/mnras/staa1128} {\bibfield  {journal}
  {\bibinfo  {journal} {Monthly Notices of the Royal Astronomical Society}\
  }\textbf {\bibinfo {volume} {494}},\ \bibinfo {pages} {6001} (\bibinfo {year}
  {2020})}\BibitemShut {NoStop}%
\bibitem [{\citenamefont {Cowan}\ \emph {et~al.}(2011)\citenamefont {Cowan},
  \citenamefont {Cranmer}, \citenamefont {Gross},\ and\ \citenamefont
  {Vitells}}]{Cowan:2010js}%
  \BibitemOpen
  \bibfield  {author} {\bibinfo {author} {\bibfnamefont {G.}~\bibnamefont
  {Cowan}}, \bibinfo {author} {\bibfnamefont {K.}~\bibnamefont {Cranmer}},
  \bibinfo {author} {\bibfnamefont {E.}~\bibnamefont {Gross}}, \ and\ \bibinfo
  {author} {\bibfnamefont {O.}~\bibnamefont {Vitells}},\ }\href {\doibase
  10.1140/epjc/s10052-011-1554-0} {\bibfield  {journal} {\bibinfo  {journal}
  {Eur. Phys. J. C}\ }\textbf {\bibinfo {volume} {71}},\ \bibinfo {pages}
  {1554} (\bibinfo {year} {2011})},\ \bibinfo {note} {[Erratum: Eur.Phys.J.C
  73, 2501 (2013)]},\ \Eprint {http://arxiv.org/abs/1007.1727} {arXiv:1007.1727
  [physics.data-an]} \BibitemShut {NoStop}%
\bibitem [{\citenamefont {{Bennett}}\ and\ \citenamefont
  {{Bovy}}(2019)}]{2019MNRAS.482.1417B}%
  \BibitemOpen
  \bibfield  {author} {\bibinfo {author} {\bibfnamefont {M.}~\bibnamefont
  {{Bennett}}}\ and\ \bibinfo {author} {\bibfnamefont {J.}~\bibnamefont
  {{Bovy}}},\ }\href {\doibase 10.1093/mnras/sty2813} {\bibfield  {journal}
  {\bibinfo  {journal} {Mon. Not. Roy. Astron. Soc.}\ }\textbf {\bibinfo
  {volume} {482}},\ \bibinfo {pages} {1417} (\bibinfo {year}
  {2019})}\BibitemShut {NoStop}%
\bibitem [{\citenamefont {Fabricius}\ \emph {et~al.}(2020)\citenamefont
  {Fabricius}, \citenamefont {Luri}, \citenamefont {Arenou}, \citenamefont
  {Babusiaux}, \citenamefont {Helmi}, \citenamefont {Muraveva}, \citenamefont
  {Reyl{\'e}}, \citenamefont {Spoto},\ and\ \citenamefont
  {Vallenari}}]{Fabricius_2020}%
  \BibitemOpen
  \bibfield  {author} {\bibinfo {author} {\bibfnamefont {C.}~\bibnamefont
  {Fabricius}}, \bibinfo {author} {\bibfnamefont {X.}~\bibnamefont {Luri}},
  \bibinfo {author} {\bibfnamefont {F.}~\bibnamefont {Arenou}}, \bibinfo
  {author} {\bibfnamefont {C.}~\bibnamefont {Babusiaux}}, \bibinfo {author}
  {\bibfnamefont {A.}~\bibnamefont {Helmi}}, \bibinfo {author} {\bibfnamefont
  {T.}~\bibnamefont {Muraveva}}, \bibinfo {author} {\bibfnamefont
  {C.}~\bibnamefont {Reyl{\'e}}}, \bibinfo {author} {\bibfnamefont
  {F.}~\bibnamefont {Spoto}}, \ and\ \bibinfo {author} {\bibfnamefont
  {A.}~\bibnamefont {Vallenari}},\ }\href {\doibase
  10.1051/0004-6361/202039834} {\bibfield  {journal} {\bibinfo  {journal}
  {Astronomy \& Astrophysics}\ } (\bibinfo {year} {2020}),\
  10.1051/0004-6361/202039834}\BibitemShut {NoStop}%
\bibitem [{\citenamefont {{Lindegren}}\ \emph {et~al.}(2020)\citenamefont
  {{Lindegren}}, \citenamefont {{Bastian}}, \citenamefont {{Biermann}},
  \citenamefont {{Bombrun}}, \citenamefont {{de Torres}}, \citenamefont
  {{Gerlach}}, \citenamefont {{Geyer}}, \citenamefont {{Hern{\'a}ndez}},
  \citenamefont {{Hilger}}, \citenamefont {{Hobbs}}, \citenamefont {{Klioner}},
  \citenamefont {{Lammers}}, \citenamefont {{McMillan}}, \citenamefont
  {{Ramos-Lerate}}, \citenamefont {{Steidelm{\"u}ller}}, \citenamefont
  {{Stephenson}},\ and\ \citenamefont {{van Leeuwen}}}]{2020arXiv201201742L}%
  \BibitemOpen
  \bibfield  {author} {\bibinfo {author} {\bibfnamefont {L.}~\bibnamefont
  {{Lindegren}}}, \bibinfo {author} {\bibfnamefont {U.}~\bibnamefont
  {{Bastian}}}, \bibinfo {author} {\bibfnamefont {M.}~\bibnamefont
  {{Biermann}}}, \bibinfo {author} {\bibfnamefont {A.}~\bibnamefont
  {{Bombrun}}}, \bibinfo {author} {\bibfnamefont {A.}~\bibnamefont {{de
  Torres}}}, \bibinfo {author} {\bibfnamefont {E.}~\bibnamefont {{Gerlach}}},
  \bibinfo {author} {\bibfnamefont {R.}~\bibnamefont {{Geyer}}}, \bibinfo
  {author} {\bibfnamefont {J.}~\bibnamefont {{Hern{\'a}ndez}}}, \bibinfo
  {author} {\bibfnamefont {T.}~\bibnamefont {{Hilger}}}, \bibinfo {author}
  {\bibfnamefont {D.}~\bibnamefont {{Hobbs}}}, \bibinfo {author} {\bibfnamefont
  {S.~A.}\ \bibnamefont {{Klioner}}}, \bibinfo {author} {\bibfnamefont
  {U.}~\bibnamefont {{Lammers}}}, \bibinfo {author} {\bibfnamefont {P.~J.}\
  \bibnamefont {{McMillan}}}, \bibinfo {author} {\bibfnamefont
  {M.}~\bibnamefont {{Ramos-Lerate}}}, \bibinfo {author} {\bibfnamefont
  {H.}~\bibnamefont {{Steidelm{\"u}ller}}}, \bibinfo {author} {\bibfnamefont
  {C.~A.}\ \bibnamefont {{Stephenson}}}, \ and\ \bibinfo {author}
  {\bibfnamefont {F.}~\bibnamefont {{van Leeuwen}}},\ }\href@noop {} {\bibfield
   {journal} {\bibinfo  {journal} {arXiv e-prints}\ ,\ \bibinfo {eid}
  {arXiv:2012.01742}} (\bibinfo {year} {2020})},\ \Eprint
  {http://arxiv.org/abs/2012.01742} {arXiv:2012.01742 [astro-ph.IM]}
  \BibitemShut {NoStop}%
\bibitem [{\citenamefont {{Vallenari}}(2018)}]{2018FrASS...5...11V}%
  \BibitemOpen
  \bibfield  {author} {\bibinfo {author} {\bibfnamefont {A.}~\bibnamefont
  {{Vallenari}}},\ }\href {\doibase 10.3389/fspas.2018.00011} {\bibfield
  {journal} {\bibinfo  {journal} {Frontiers in Astronomy and Space Sciences}\
  }\textbf {\bibinfo {volume} {5}},\ \bibinfo {eid} {11} (\bibinfo {year}
  {2018})}\BibitemShut {NoStop}%
\bibitem [{\citenamefont {{The Theia Collaboration}}\ \emph
  {et~al.}(2017)\citenamefont {{The Theia Collaboration}}, \citenamefont
  {{Boehm}}, \citenamefont {{Krone-Martins}}, \citenamefont {{Amorim}},
  \citenamefont {{Anglada-Escude}}, \citenamefont {{Brandeker}}, \citenamefont
  {{Courbin}}, \citenamefont {{Ensslin}}, \citenamefont {{Falcao}},
  \citenamefont {{Freese}}, \citenamefont {{Holl}}, \citenamefont {{Labadie}},
  \citenamefont {{Leger}}, \citenamefont {{Malbet}}, \citenamefont {{Mamon}}
  \emph {et~al.}}]{2017arXiv170701348T}%
  \BibitemOpen
  \bibfield  {author} {\bibinfo {author} {\bibnamefont {{The Theia
  Collaboration}}}, \bibinfo {author} {\bibfnamefont {C.}~\bibnamefont
  {{Boehm}}}, \bibinfo {author} {\bibfnamefont {A.}~\bibnamefont
  {{Krone-Martins}}}, \bibinfo {author} {\bibfnamefont {A.}~\bibnamefont
  {{Amorim}}}, \bibinfo {author} {\bibfnamefont {G.}~\bibnamefont
  {{Anglada-Escude}}}, \bibinfo {author} {\bibfnamefont {A.}~\bibnamefont
  {{Brandeker}}}, \bibinfo {author} {\bibfnamefont {F.}~\bibnamefont
  {{Courbin}}}, \bibinfo {author} {\bibfnamefont {T.}~\bibnamefont
  {{Ensslin}}}, \bibinfo {author} {\bibfnamefont {A.}~\bibnamefont {{Falcao}}},
  \bibinfo {author} {\bibfnamefont {K.}~\bibnamefont {{Freese}}}, \bibinfo
  {author} {\bibfnamefont {B.}~\bibnamefont {{Holl}}}, \bibinfo {author}
  {\bibfnamefont {L.}~\bibnamefont {{Labadie}}}, \bibinfo {author}
  {\bibfnamefont {A.}~\bibnamefont {{Leger}}}, \bibinfo {author} {\bibfnamefont
  {F.}~\bibnamefont {{Malbet}}}, \bibinfo {author} {\bibfnamefont
  {G.}~\bibnamefont {{Mamon}}},  \emph {et~al.},\ }\href@noop {} {\bibfield
  {journal} {\bibinfo  {journal} {arXiv e-prints}\ } (\bibinfo {year}
  {2017})},\ \Eprint {http://arxiv.org/abs/1707.01348} {arXiv:1707.01348
  [astro-ph.IM]} \BibitemShut {NoStop}%
\bibitem [{\citenamefont {{Hobbs}}\ \emph {et~al.}(2016)\citenamefont
  {{Hobbs}}, \citenamefont {{H{\o}g}}, \citenamefont {{Mora}}, \citenamefont
  {{Crowley}}, \citenamefont {{McMillan}}, \citenamefont {{Ranalli}},
  \citenamefont {{Heiter}}, \citenamefont {{Jordi}}, \citenamefont {{Hambly}},
  \citenamefont {{Church}}, \citenamefont {{Anthony}}, \citenamefont {{Tanga}},
  \citenamefont {{Chemin}}, \citenamefont {{Portell}}, \citenamefont
  {{Jim{\'e}nez-Esteban}} \emph {et~al.}}]{2016arXiv160907325H}%
  \BibitemOpen
  \bibfield  {author} {\bibinfo {author} {\bibfnamefont {D.}~\bibnamefont
  {{Hobbs}}}, \bibinfo {author} {\bibfnamefont {E.}~\bibnamefont {{H{\o}g}}},
  \bibinfo {author} {\bibfnamefont {A.}~\bibnamefont {{Mora}}}, \bibinfo
  {author} {\bibfnamefont {C.}~\bibnamefont {{Crowley}}}, \bibinfo {author}
  {\bibfnamefont {P.}~\bibnamefont {{McMillan}}}, \bibinfo {author}
  {\bibfnamefont {P.}~\bibnamefont {{Ranalli}}}, \bibinfo {author}
  {\bibfnamefont {U.}~\bibnamefont {{Heiter}}}, \bibinfo {author}
  {\bibfnamefont {C.}~\bibnamefont {{Jordi}}}, \bibinfo {author} {\bibfnamefont
  {N.}~\bibnamefont {{Hambly}}}, \bibinfo {author} {\bibfnamefont
  {R.}~\bibnamefont {{Church}}}, \bibinfo {author} {\bibfnamefont
  {B.}~\bibnamefont {{Anthony}}}, \bibinfo {author} {\bibfnamefont
  {P.}~\bibnamefont {{Tanga}}}, \bibinfo {author} {\bibfnamefont
  {L.}~\bibnamefont {{Chemin}}}, \bibinfo {author} {\bibfnamefont
  {J.}~\bibnamefont {{Portell}}}, \bibinfo {author} {\bibfnamefont
  {F.}~\bibnamefont {{Jim{\'e}nez-Esteban}}},  \emph {et~al.},\ }\href@noop {}
  {\bibfield  {journal} {\bibinfo  {journal} {arXiv e-prints}\ ,\ \bibinfo
  {eid} {arXiv:1609.07325}} (\bibinfo {year} {2016})},\ \Eprint
  {http://arxiv.org/abs/1609.07325} {arXiv:1609.07325 [astro-ph.IM]}
  \BibitemShut {NoStop}%
\bibitem [{\citenamefont {Mishra-Sharma}\ \emph {et~al.}(2020)\citenamefont
  {Mishra-Sharma}, \citenamefont {Van~Tilburg},\ and\ \citenamefont
  {Weiner}}]{Mishra-Sharma:2020ynk}%
  \BibitemOpen
  \bibfield  {author} {\bibinfo {author} {\bibfnamefont {S.}~\bibnamefont
  {Mishra-Sharma}}, \bibinfo {author} {\bibfnamefont {K.}~\bibnamefont
  {Van~Tilburg}}, \ and\ \bibinfo {author} {\bibfnamefont {N.}~\bibnamefont
  {Weiner}},\ }\href {\doibase 10.1103/PhysRevD.102.023026} {\bibfield
  {journal} {\bibinfo  {journal} {Phys. Rev. D}\ }\textbf {\bibinfo {volume}
  {102}},\ \bibinfo {pages} {023026} (\bibinfo {year} {2020})},\ \Eprint
  {http://arxiv.org/abs/2003.02264} {arXiv:2003.02264 [astro-ph.CO]}
  \BibitemShut {NoStop}%
\bibitem [{\citenamefont {Mondino}\ \emph {et~al.}(2020)\citenamefont
  {Mondino}, \citenamefont {Taki}, \citenamefont {Van~Tilburg},\ and\
  \citenamefont {Weiner}}]{Mondino:2020rkn}%
  \BibitemOpen
  \bibfield  {author} {\bibinfo {author} {\bibfnamefont {C.}~\bibnamefont
  {Mondino}}, \bibinfo {author} {\bibfnamefont {A.-M.}\ \bibnamefont {Taki}},
  \bibinfo {author} {\bibfnamefont {K.}~\bibnamefont {Van~Tilburg}}, \ and\
  \bibinfo {author} {\bibfnamefont {N.}~\bibnamefont {Weiner}},\ }\href
  {\doibase 10.1103/PhysRevLett.125.111101} {\bibfield  {journal} {\bibinfo
  {journal} {Phys. Rev. Lett.}\ }\textbf {\bibinfo {volume} {125}},\ \bibinfo
  {pages} {111101} (\bibinfo {year} {2020})},\ \Eprint
  {http://arxiv.org/abs/2002.01938} {arXiv:2002.01938 [astro-ph.CO]}
  \BibitemShut {NoStop}%
\bibitem [{\citenamefont {Fan}\ \emph {et~al.}(2013{\natexlab{a}})\citenamefont
  {Fan}, \citenamefont {Katz}, \citenamefont {Randall},\ and\ \citenamefont
  {Reece}}]{Fan:2013tia}%
  \BibitemOpen
  \bibfield  {author} {\bibinfo {author} {\bibfnamefont {J.}~\bibnamefont
  {Fan}}, \bibinfo {author} {\bibfnamefont {A.}~\bibnamefont {Katz}}, \bibinfo
  {author} {\bibfnamefont {L.}~\bibnamefont {Randall}}, \ and\ \bibinfo
  {author} {\bibfnamefont {M.}~\bibnamefont {Reece}},\ }\href {\doibase
  10.1103/PhysRevLett.110.211302} {\bibfield  {journal} {\bibinfo  {journal}
  {Phys. Rev. Lett.}\ }\textbf {\bibinfo {volume} {110}},\ \bibinfo {pages}
  {211302} (\bibinfo {year} {2013}{\natexlab{a}})},\ \Eprint
  {http://arxiv.org/abs/1303.3271} {arXiv:1303.3271 [hep-ph]} \BibitemShut
  {NoStop}%
\bibitem [{\citenamefont {Fan}\ \emph {et~al.}(2013{\natexlab{b}})\citenamefont
  {Fan}, \citenamefont {Katz}, \citenamefont {Randall},\ and\ \citenamefont
  {Reece}}]{Fan:2013yva}%
  \BibitemOpen
  \bibfield  {author} {\bibinfo {author} {\bibfnamefont {J.}~\bibnamefont
  {Fan}}, \bibinfo {author} {\bibfnamefont {A.}~\bibnamefont {Katz}}, \bibinfo
  {author} {\bibfnamefont {L.}~\bibnamefont {Randall}}, \ and\ \bibinfo
  {author} {\bibfnamefont {M.}~\bibnamefont {Reece}},\ }\href {\doibase
  10.1016/j.dark.2013.07.001} {\bibfield  {journal} {\bibinfo  {journal} {Phys.
  Dark Univ.}\ }\textbf {\bibinfo {volume} {2}},\ \bibinfo {pages} {139}
  (\bibinfo {year} {2013}{\natexlab{b}})},\ \Eprint
  {http://arxiv.org/abs/1303.1521} {arXiv:1303.1521 [astro-ph.CO]} \BibitemShut
  {NoStop}%
\bibitem [{\citenamefont {Navarro}\ \emph {et~al.}(1996)\citenamefont
  {Navarro}, \citenamefont {Frenk},\ and\ \citenamefont
  {White}}]{Navarro:1995iw}%
  \BibitemOpen
  \bibfield  {author} {\bibinfo {author} {\bibfnamefont {J.~F.}\ \bibnamefont
  {Navarro}}, \bibinfo {author} {\bibfnamefont {C.~S.}\ \bibnamefont {Frenk}},
  \ and\ \bibinfo {author} {\bibfnamefont {S.~D.~M.}\ \bibnamefont {White}},\
  }\href {\doibase 10.1086/177173} {\bibfield  {journal} {\bibinfo  {journal}
  {Astrophys. J.}\ }\textbf {\bibinfo {volume} {462}},\ \bibinfo {pages} {563}
  (\bibinfo {year} {1996})},\ \Eprint {http://arxiv.org/abs/astro-ph/9508025}
  {astro-ph/9508025} \BibitemShut {NoStop}%
\bibitem [{\citenamefont {Navarro}\ \emph {et~al.}(1997)\citenamefont
  {Navarro}, \citenamefont {Frenk},\ and\ \citenamefont
  {White}}]{Navarro:1996gj}%
  \BibitemOpen
  \bibfield  {author} {\bibinfo {author} {\bibfnamefont {J.~F.}\ \bibnamefont
  {Navarro}}, \bibinfo {author} {\bibfnamefont {C.~S.}\ \bibnamefont {Frenk}},
  \ and\ \bibinfo {author} {\bibfnamefont {S.~D.~M.}\ \bibnamefont {White}},\
  }\href {\doibase 10.1086/304888} {\bibfield  {journal} {\bibinfo  {journal}
  {Astrophys. J.}\ }\textbf {\bibinfo {volume} {490}},\ \bibinfo {pages} {493}
  (\bibinfo {year} {1997})},\ \Eprint {http://arxiv.org/abs/astro-ph/9611107}
  {arXiv:astro-ph/9611107 [astro-ph]} \BibitemShut {NoStop}%
\bibitem [{\citenamefont {Aad}\ \emph {et~al.}(2014)\citenamefont {Aad} \emph
  {et~al.}}]{Aad:2014eha}%
  \BibitemOpen
  \bibfield  {author} {\bibinfo {author} {\bibfnamefont {G.}~\bibnamefont
  {Aad}} \emph {et~al.} (\bibinfo {collaboration} {ATLAS}),\ }\href {\doibase
  10.1103/PhysRevD.90.112015} {\bibfield  {journal} {\bibinfo  {journal} {Phys.
  Rev. D}\ }\textbf {\bibinfo {volume} {90}},\ \bibinfo {pages} {112015}
  (\bibinfo {year} {2014})},\ \Eprint {http://arxiv.org/abs/1408.7084}
  {arXiv:1408.7084 [hep-ex]} \BibitemShut {NoStop}%
\end{thebibliography}%

\clearpage

\clearpage

\onecolumngrid
\begin{center}
  \textbf{\large Supplemental Material for: The Galactic potential and dark matter density from angular stellar accelerations}\\[.2cm]
  \vspace{0.05in}
  {Malte Buschmann, \ Benjamin R. Safdi, \ and \ Katelin Schutz}
\end{center}

This Supplemental Material (SM) is organized as follows. In Sec.~\ref{app:Pot} we give additional formulae describing the potential models that we use in the analyses presented in the main {\it Letter}. In Sec.~\ref{app:analytic} we present additional steps leading up to the derivation of~\eqref{eq:delta_chi2_scale} in the main body. Sec.~\ref{app:GH} gives addition details behind the analysis of {\it Gaia}-{\it Hipparcos} data, while Sec.~\ref{app:extended} presents additional results from the {\it Gaia} projections discussed in the main body. 

\section{Potential model components}
\label{app:Pot}

In the following analyses and projections we include three components for the Galactic potential. These components are simplified and account for the Galactic disk ($\Phi_{\rm disk}$), the Galactic bulge ($\Phi_{\rm bulge}$), and the DM halo ($\Phi_{\rm DM}$).
Recall that the stellar accelerations are then given by ${\bf a}^{\rm GF} = - \nabla \Phi_{\rm tot}({\bf x})$.

In this work we consider (i) the ability to detect the total gravitational potential through stellar accelerations, and then (ii) after detecting the total potential the ability to discriminate its contributions, and the contribution from the Galactic DM in particular. For the first step, it is useful to define a scaling constant $\lambda$ such that $\Phi_{\rm tot}^\lambda({\bf x}) = \lambda \times \Phi_{\rm tot}({\bf x})$, with $\Phi_{\rm tot}({\bf x})$ being a representative Galactic potential model that has fixed model parameters, so that the only free model parameter to be determined by fitting to the data is the scaling parameter $\lambda$ (with expected value $\lambda = 1$). Note that $\lambda$ is equivalent to a parameter that scales the total mass of the Milky Way. When addressing the second aim we will instead consider model parameters that individually adjust, for example, the disk, bulge, and DM halo masses.

\subsection{The Galactic disk}

We make use of the Miyamoto-Nagai disk model~\cite{1975PASJ...27..533M} to account for the gravitational potential of the disk. The gravitational potential for this model is 
\es{}{
\Phi_{\rm disk}(R,z) = - {G M_D \over \sqrt{R^2 + \left[ a + \sqrt{z^2 + b^2} \right]^2}} \,,}
where $M_D$ is the disk mass, $a$ and $b$ are the scale radius and scale height of the disk, and $R$ ($z$) is the cylindrical radial (vertical) coordinate. We base the truth values on the best-fit values for the Milky Way from~\cite{2013ApJ...779..115B,2015ApJS..216...29B}, which found $M_D = 6.8 \times 10^{10}$ $M_\odot$, $a = 3.0$ kpc, and $b = 280$ pc.

\subsection{The Galactic Bulge}
The form of the Galactic bulge density profile does not play a significant role in this work because most of stars are located outside of the Inner Galaxy, so that the bulge may simply be replaced by a point mass at the Galactic Center with no loss of accuracy. However, for completeness we model the bulge as in~\cite{2013ApJ...779..115B} and take the density profile to be a Hernquist sphere, 
\es{}{
\Phi_{\rm bulge}(r) = - \frac{G M_b}{r+a},
}
where $a$ is the scale radius, which we take to be 0.6 kpc~\cite{2013ApJ...779..115B}. Ref.~\cite{2015ApJS..216...29B} finds the best-fit bulge mass of $M_b = 5 \times 10^9$ $M_\odot$, and we use this mass as our default truth value when making projections below. 

\subsection{Dark Matter}

We model the DM density profile as a 
NFW spherically-symmetric density profile~\cite{Navarro:1995iw,Navarro:1996gj}. The NFW DM profile is given by 
\es{NFW}{
\rho_\text{DM}(r) = {\rho_0 \over {r / r_s} \left( 1 + {r / r_s} \right)^2} \,,}
where $r_s$ is the scale radius, and $\rho_0$ is the characteristic density that may be related to the local DM density at $r_\odot \approx 8.22$ kpc~\cite{bovy2020purely}. As our fiducial truth scenario we again adopt the best-fit values presented in~\cite{2015ApJS..216...29B}, with the local DM density $\rho_\odot = 0.008$ $M_\odot$/pc$^3$ and $r_s = 16$ kpc.

\section{Analytic likelihood approximation: extended results}
\label{app:analytic}

Here we fill in additional steps leading up to the result in~\eqref{eq:delta_chi2_scale}, which assumes that all stars are in the Galactic plane. We choose a coordinate system where the Sun is located at $\vec{r}_\odot = r_\odot {\bf \hat x}$. In the Galactic frame the solar acceleration is given by
\es{}{
{\bf a}_\odot = -a(r_\odot) {\bf \hat x} \,,}
where $a(r_\odot)$ is the magnitude of the acceleration in the radial direction. A star a distance $r_i$ from the Sun at Galactic longitude $\ell_i$ then feels acceleration
\es{}{
{\bf a}^{\rm GF}(r_i)= a\left(\sqrt{r_i^2 + r_\odot^2 -2 r_i r_\odot \cos\ell_i}\right) {\left[ (r_i \cos \ell_i - r_\odot) {\bf \hat x} + r_i\sin \ell_i {\bf \hat y} \right] \over \sqrt{r_i^2 + r_\odot^2 -2 r_i r_\odot \cos\ell_i} } \,.
}
For $r_i / r_\odot \ll 1$, the solar-frame velocities may be approximated by
\es{}{
{\bf a}(r_i) &\equiv {\bf a}^{\rm GF}(r_i) - {\bf a}_\odot \\
&\approx r_i \left( a'(r_\odot) \cos \ell_i {\bf \hat x} + {a(r_\odot) \sin \ell_i \over r_\odot} {\bf \hat y} \right)  + \mathcal{O}\left({r_i \over r_\odot}\right)^2 \,.
}
To project this motion onto the transverse direction relative to the Sun, we define the unit vector
${\bf \hat T}_i$, which points in the direction of the PM:
\es{}{
{\bf \hat T}_i = \sin \ell_i {\bf \hat x} - \cos \ell_i {\bf \hat y} \,.
}
Thus the component of the acceleration in the direction of the PM is
\es{}{
{\bf a}_i^{T} \approx {r_i \over r_\odot} \cos \ell_i \sin \ell_i \left[- a(r_\odot) + r_\odot a'(r_\odot) \right] \,.
}
The PM accelerations are $\alpha_i = \varpi_i {\bf a}_i^{T}$, with $\varpi_i = 1/ r_i$ the parallax. Thus the difference of test statistic between the signal model and the null model under the Asimov signal hypothesis is
\es{eq:delta_chi2_scale-detail}{
q&\approx {N \over 8 \sigma_{\alpha_T}^2} {\left[-a(r_\odot) + r_\odot a'(r_\odot) \right]^2 \over r_\odot^2 }\,   \\
 &\approx {N \over 8 \sigma_{\alpha_T}^2} {\left[3 \, {a(r_\odot) \over r_\odot} + 4 \pi G \bar \rho(r_\odot) \right]^2   }
\,,
}
where in the last step we have used the relation $a'(r_\odot) \approx - 2 a(r_\odot) / r_\odot - 4 \pi G \bar \rho(r_\odot)$, and where $\bar \rho(r_\odot) \equiv {1 \over 4 \pi} \int d\Omega \rho(r_\odot, \Omega)$ is the matter density at Galactic radius $r_\odot$ averaged over a spherical shell surrounding the Galactic Center.

Let us now consider the model parameter $\theta_1 \equiv a(r_\odot) / r_\odot$ and $\theta_2 \equiv \bar \rho(r_\odot)$ separately. We compute the Fisher matrix using the Asimov formalism assuming true value $\hat \theta_1$ and $\hat \theta_2$ with deformation ${\bm \delta \theta} = (\delta \theta_1,\delta \theta_2)$, leading to $\delta \chi^2 = {\bm \delta \theta}^T \cdot {\bm I} \cdot {\bm \delta \theta}$, with ${\bm I}$ the Fisher matrix. Inverting the Fisher matrix yields the expected covariance matrix ${\bm \Sigma}$:
\es{}{
{\bm \Sigma} ={4 \over 3} {\sigma_{\alpha_T}^2 \over N} \left(
\begin{array}{cc}
{8 \over 9} & -{1 \over 3}  {1 \over \pi G}  \\
-{1 \over 3}  {1 \over \pi G}  & {1 \over 2} {1 \over (\pi G)^2}
\end{array}
\right) \,, 
}
which implies, for example, that the expected uncertainty on the DM density, profiling over $\theta_1$, is  
\es{}{
{\sigma_{\rho_{\rm DM}} \over \rho_{\rm DM} } = \sqrt{2 \over 3} {\sigma_{\alpha_T}\over \sqrt{N}  \pi G \rho_{\rm DM} }  \approx 2.9 \left( { 10 \, \, {\rm yr} \over T} \right)^{5/2} \,,
}
while similarly
\es{}{
{ \sigma_{a_\odot / r_\odot} \over a_\odot / r_\odot } = {4 \over 3} \sqrt{2 \over 3} {\sigma_{\alpha_T}\over \sqrt{N}  |a(r_\odot)| / r_\odot }  \approx 0.49 \left( { 10 \, \, {\rm yr} \over T} \right)^{5/2} \,,
}
Thus, with 10 yrs of {\it Gaia} data we expect, using the in-plane motion, to measure the Galactic acceleration at $\sim$2$\sigma$ but not quite have the sensitivity to measure the DM density independently. With that said, as shown in the main {\it Letter}, if additional prior information is incorporated that informs $a(r_\odot)$, then the 10-year {\it Gaia} data set may be sensitive to $\rho_{\rm DM}$ at more than 1$\sigma$.

\section{{\it Gaia}-{\it Hipparcos} analysis}
\label{app:GH}

In this section we present additional details behind the analysis of the joint {\it Gaia} DR2 -- {\it Hipparcos} acceleration data. We begin with a summary of the data processing that we perform to extract the acceleration data, and then we describe additional analysis details and cross checks.

\subsection{Joining Hipparos and {\it Gaia} for acceleration measurements}
\label{sec:GH_accel}

In the absence of stellar acceleration measurements determined from the {\it Gaia} data alone we combine the {\it Hipparcos} and {\it Gaia} DR2 astrometric data sets to obtain estimates of the {\it Hipparcos} star accelerations, taking advantage of the $\sim$24 yr baseline between the data sets. Such acceleration measurements were already performed and cataloged in~\cite{2018ApJS..239...31B}, though there are minor processing steps needed to convert the data into the format needed for our analysis. In particular, Ref.~\cite{2018ApJS..239...31B} compared the Hipparcos and {\it Gaia} stellar positions to obtain a PM measure: $\mu^{GH}_\alpha = (\alpha_G - \alpha_H)/T_{GH}$, where $\alpha_G$ ($\alpha_H$) is the RA measurement from {\it Gaia} (Hipparcos) and $T_{GH} \approx 24.25$ yr is the baseline between the two reference epochs for the two surveys. A similar expression holds for the DEC PM $\mu_\delta^{GH}$. The key point is that an astrometric solution with accelerations would predict {\it e.g.} $\mu_\alpha^{GH} = \mu_\alpha + {1 \over 2} \gamma_\alpha T_{GH}$ and a {\it Gaia}-only PM of $\mu_\alpha^{G} = \mu_\alpha + \gamma_\alpha T_{GH}$, where the superscript denotes measurements made at the {\it Gaia} epoch and by definition $\mu_\alpha$ being the PM at the {\it Hipparcos} epoch. We may thus use the difference between $\mu_\alpha^G$ and $\mu_{\alpha}^{GH}$ to infer the acceleration $\gamma_\alpha$ (and similarly for $\gamma_\delta$).

In practice, we do not just want the central values of the accelerations $\gamma_\alpha$ and $\gamma_\delta$, but we also want to compute the associated covariance matrix. Towards that end, for each star in the {\it Gaia}-{\it Hipparcos} catalog we construct the loss function
\es{eq:chi2_GH}{
\chi^2(\mu_\alpha, \gamma_\alpha, \mu_\delta, \gamma_\delta) = &{\bf x}_G \cdot {\bf C}_G^{-1} \cdot {\bf x}_G + {\bf x}_{GH} \cdot {\bf C}_{GH}^{-1} \cdot {\bf x}_{GH}  \,,
}
where 
\es{}{
{\bf x}_G = \left( {\mu_\alpha^G - \mu_\alpha^{0,G} \over \sigma_{\mu_\alpha^G} } ,{\mu_\delta^G - \mu_\delta^{0,G} \over \sigma_{\mu_\delta^G} } \right) \,,
}
and the superscript $0$ denotes a measured quantity. A similar expression holds for the ${\bf x}_{GH}$ vector. Note that the correlation matrices ${\bf C}_G$ and ${\bf C}_{GH}$ are provided in the {\it Gaia} and {\it Gaia}-{\it Hipparcos} catalogs, respectively. We find the best-fit values for the PM velocities and accelerations, along with the associated covariance matrices, by minimizing the loss function in~\eqref{eq:chi2_GH} and computing the Hessian matrix. 

 \begin{figure*}[htb]
\begin{center}
\includegraphics[width = .49\textwidth]{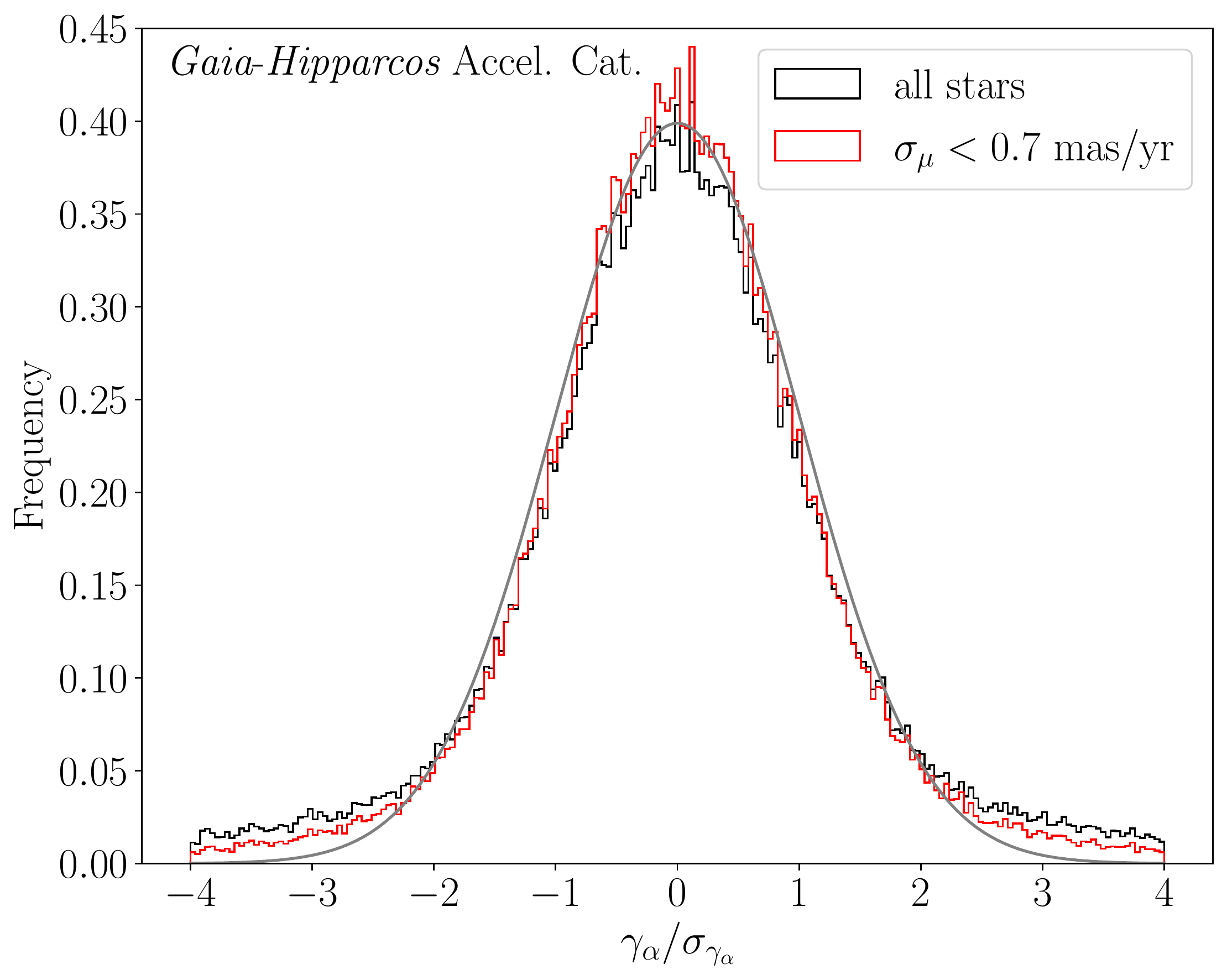}
\includegraphics[width = .49\textwidth]{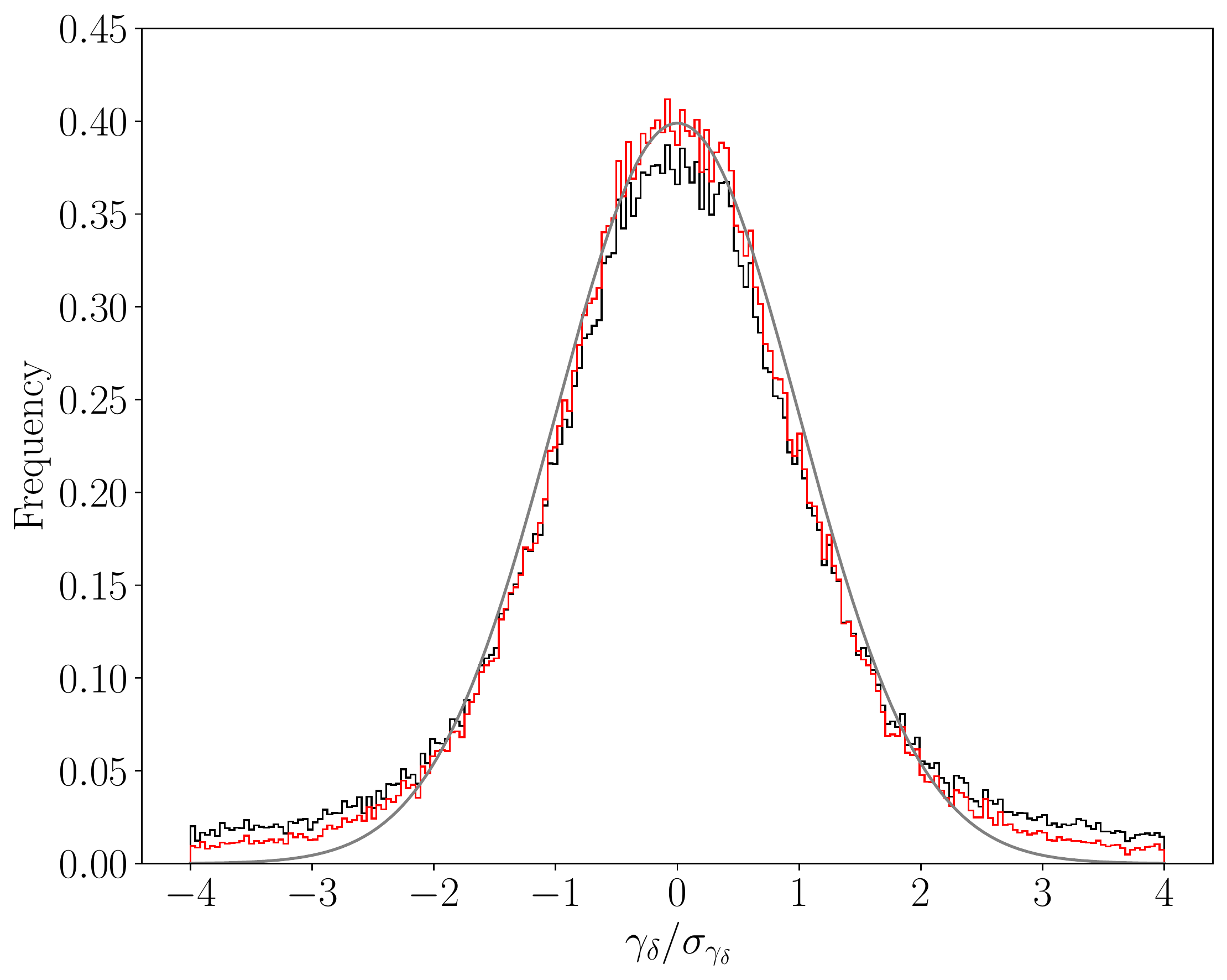}
\end{center}
\caption{(Left) A normalized histogram of the RA accelerations, relative to their uncertainties, in the {\it Gaia}-Hipparcos acceleration catalog. In both black and red we exclude stars that are over four standard deviations away from zero in acceleration, but in red we additionally include a cut on the {\it Gaia} PM uncertainties, which brings the data closer in line with the Gaussian expectation, indicated in grey. (Right) As in the left panel but for the DEC accelerations. } 
\label{fig:GH_data}
\end{figure*}

In Fig.~\ref{fig:GH_data} we illustrate the acceleration data found from this procedure. In the left (right) panel we histogram the values of $\gamma_\alpha$ ($\gamma_\delta$) relative to their uncertainties $\sigma_{\gamma_\alpha}$ ($\sigma_{\gamma_\delta}$). If the uncertainties were normally distributed then these distributions would follow a normal distribution with variance of unity, which is illustrated in light grey. The distributions are seen to be slightly wider, with extended tails. As noted in~\cite{2018ApJS..239...31B} part of the reason for the apparent non-Gaussianity is due to bright stars, whose PMs are not well measured by {\it Gaia}. We follow the recommended quality cut in~\cite{2018ApJS..239...31B} and remove stars with {\it Gaia} PM uncertainties less than 0.7 mas/yr, leading to the red curves. 

As mentioned in the main text, we apply additional quality cuts to the {\it Gaia}-{\it Hipparcos} data. First, we remove stars for which either $| \gamma_\alpha / \sigma_{\gamma_\alpha} | > 4$ or $| \gamma_\delta / \sigma_{\gamma_\delta} | > 4$ in order to remove outlier stars. From MC simulations this cut should have no noticeable effect on either the null or signal hypotheses. Second, we remove stars whose parallax measurement places them more than 5 kpc away from the Earth in order to remove stars with anomalous parallax measurements. Similarly, we do not include stars with negative parallax.

\subsection{Analysis results}

As discussed in the main {\it Letter}, we analyze the {\it Gaia}-{\it Hipparcos} acceleration data using the joint likelihood:
\es{eq:j-LL}{
p({\bf d}|{\bm \theta}) = \prod_{i=1}^N p_i({\bf d}_i | {\bm \theta}) \,,
}
with $p_i({\bf d}_i | {\bm \theta})$ defined for an individual star, indexed by $i$, in~\eqref{eq:p_i_deff}. It is worth commenting briefly on why in~\eqref{eq:p_i_deff} we do not include contributions from the parallax and position uncertainties. In general, these uncertainties may be included and the associated nuisance parameters may be profiled over. Given that the astrometric solution is assumed to have normally-distributed errors, as described by the covariance matrix, profiling over the other nuisance parameters ({\it e.g.}, a nuisance parameter describing the stellar parallax) may even be done analytically, since the likelihood has a simple Gaussian form. Thus accounting for parallax and position uncertainties is almost computationally equivalent to ignoring them. However, the reason we chose to ignore these uncertainties in our examples is that we expect them to be far sub-dominant compared to the acceleration uncertainties. More precisely, the parallax uncertainty $\sigma_{\varpi_i}$ induces an effective uncertainty $\delta \gamma_i$ on the model prediction for the angular acceleration $\gamma_i$ for star $i$ on the order $\delta \gamma_i \sim \gamma_i \sigma_{ \varpi_i}/\varpi_i$. Given that the parallax uncertainties obey $\sigma_{\varpi_i}/\varpi_i \lesssim 1$, it is thus clear that the uncertainties induced by parallax on the acceleration model prediction are far sub-dominant compared the measurement uncertainties on the angular accelerations themselves, which always satisfy $\sigma_{\gamma_i} /\gamma_i \gg 1$ for an individual star, with $\sigma_{\gamma_i}$ denoting the measured angular acceleration uncertainty. The same logic applies to the position uncertainties, though in that case the effect is even smaller since the positions are measured to higher precision than parallax. In summary, the angular acceleration uncertainties dominate because they are extremely large relative to the model prediction for an individual star, whereas the other astrometric parameters are measured at superior relative precision. On the other hand, if there are correlations between stars' parallax and position uncertainties, then these correlated uncertainties may be important to include, since they would no longer average down with including the ensemble of stars in the joint likelihood. Similarly, it is important to verify that the quadratic model prediction for the angular position, specified by a constant PM and acceleration, is not breaking down for a large number of the included stars, which may be the case if, for example, an analysis focuses only on stars extremely close to the Galactic Center, where a full solution to the equations of motion is needed. 

\begin{figure}[t]
\begin{center}
\includegraphics[width = .55\textwidth]{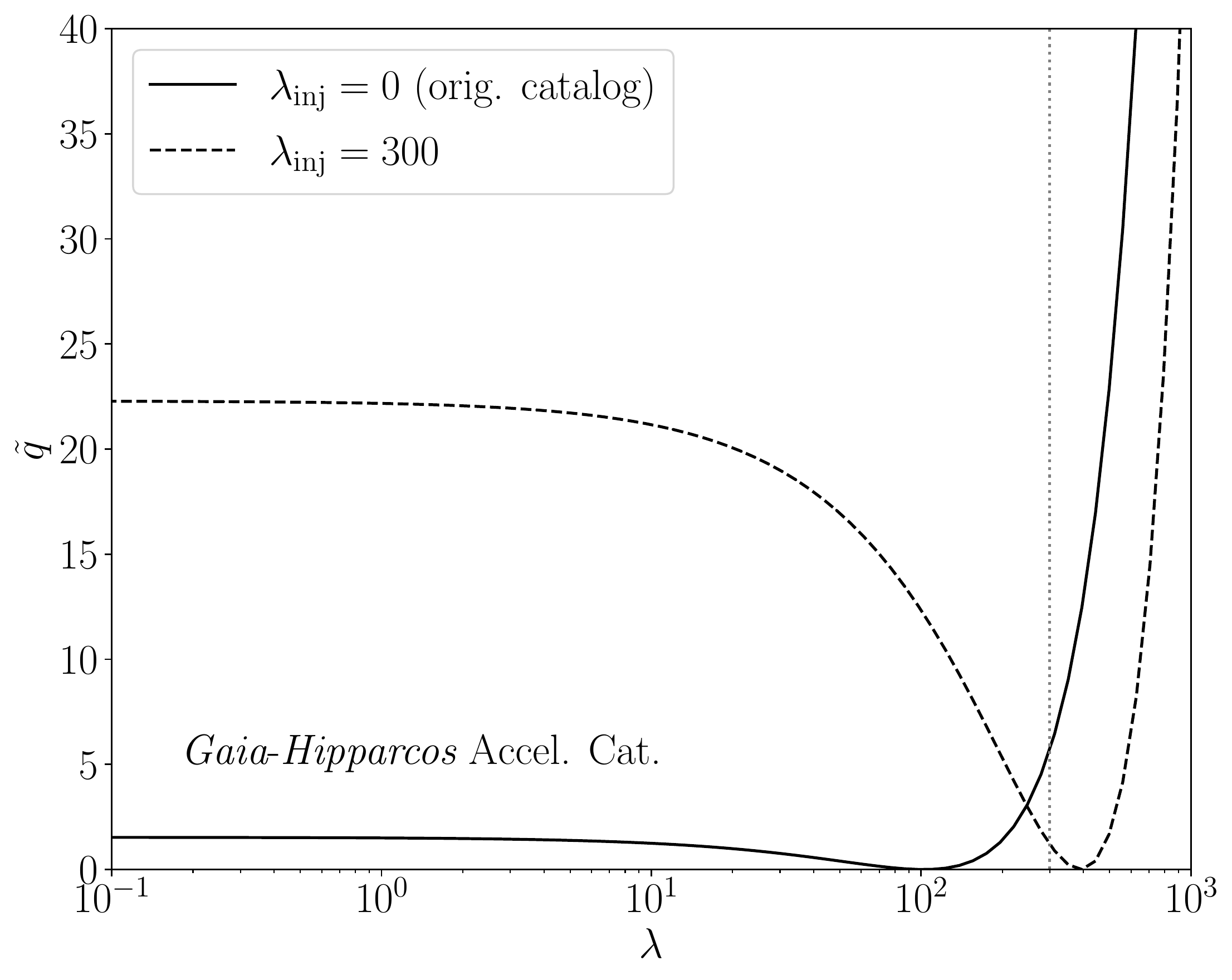}
\end{center}
\caption{The profile likelihood $\tilde q$, defined in~\eqref{eq:prof_LL}, as a function of the one-parameter-model MW potential normalization $\lambda$, for the analysis of the {\it Gaia}-{\it Hipparcos} data set. The expected value is $\lambda \approx 1$ for the actual potential, indicating that future surveys need to improve in sensitivity to $\lambda$ by around two orders of magnitude in order to discover the Galactic potential with the acceleration approach. We also illustrate the profile likelihood in the case where a synthetic signal with $\lambda_{\rm inj} = 300$ is added to the real data. In this case the analysis of the hybrid data correctly recovers the injected $\lambda$, denoted by the vertical line, to within $\sim$1$\sigma$. } 
\label{fig:GH_chi2}
\end{figure}

After quality cuts the {\it Gaia}-{\it Hipparcos} catalog has $N = 86,201$ stars. We analyze the likelihood in the context of the one-parameter model, ${\bm \theta} = \{ \lambda \}$, where $\lambda$ re-scales the full potential. The parameter $\lambda$ is physically only well defined for positive values, but for statistical consistency we analytically extend the model predictions to negative $\lambda$, since it is possible that the best-fit value for $\lambda$ would be at negative values. We then construct the profile likelihood for the parameter $\lambda$:
\es{eq:prof_LL}{
\tilde q(\lambda) = 2 \left(\log p({\bf d}| \hat { \lambda}) - \log p({\bf d}| \lambda)  \right) \,,  
}
where $\hat \lambda$ is the value that maximizes the likelihood. By definition $\tilde q(\lambda = 0) = q$, with $q$ the discovery test statistic, if $\hat \lambda > 0$, otherwise $q = 0$~\cite{Cowan:2010js}. That is, the discovery test statistic is set to zero for non-physical best-fit values. Under the null hypothesis we thus expect $q$ to follow the the one-sided chi-square distribution with one degree of freedom. In the main text we quoted the best-fit value $\hat \lambda \approx 10^2$ with $q \approx 1.5$. In Fig.~\ref{fig:GH_chi2} we show the profile likelihood over a range $\lambda$. Note that the one-sided 95\% upper limit is given by the value $\lambda^{95} > \hat \lambda$ with $\tilde q(\lambda^{95}) \approx 2.71$~\cite{Cowan:2010js}.

\subsection{Injected signal}

As a consistency check of the analysis framework we inject a synthetic signal into the {\it Gaia}-{\it Hipparcos} acceleration catalog and we verify our ability to correctly reconstruct the signal. Specifically, we inject a value $\lambda_{\rm inj} = 300$, meaning that we calculate the predicted accelerations for this value of $\lambda$ and then add to each stellar acceleration the appropriately-predicted acceleration boost. 

\begin{figure}[htb]
\begin{center}
\includegraphics[width = .49\textwidth]{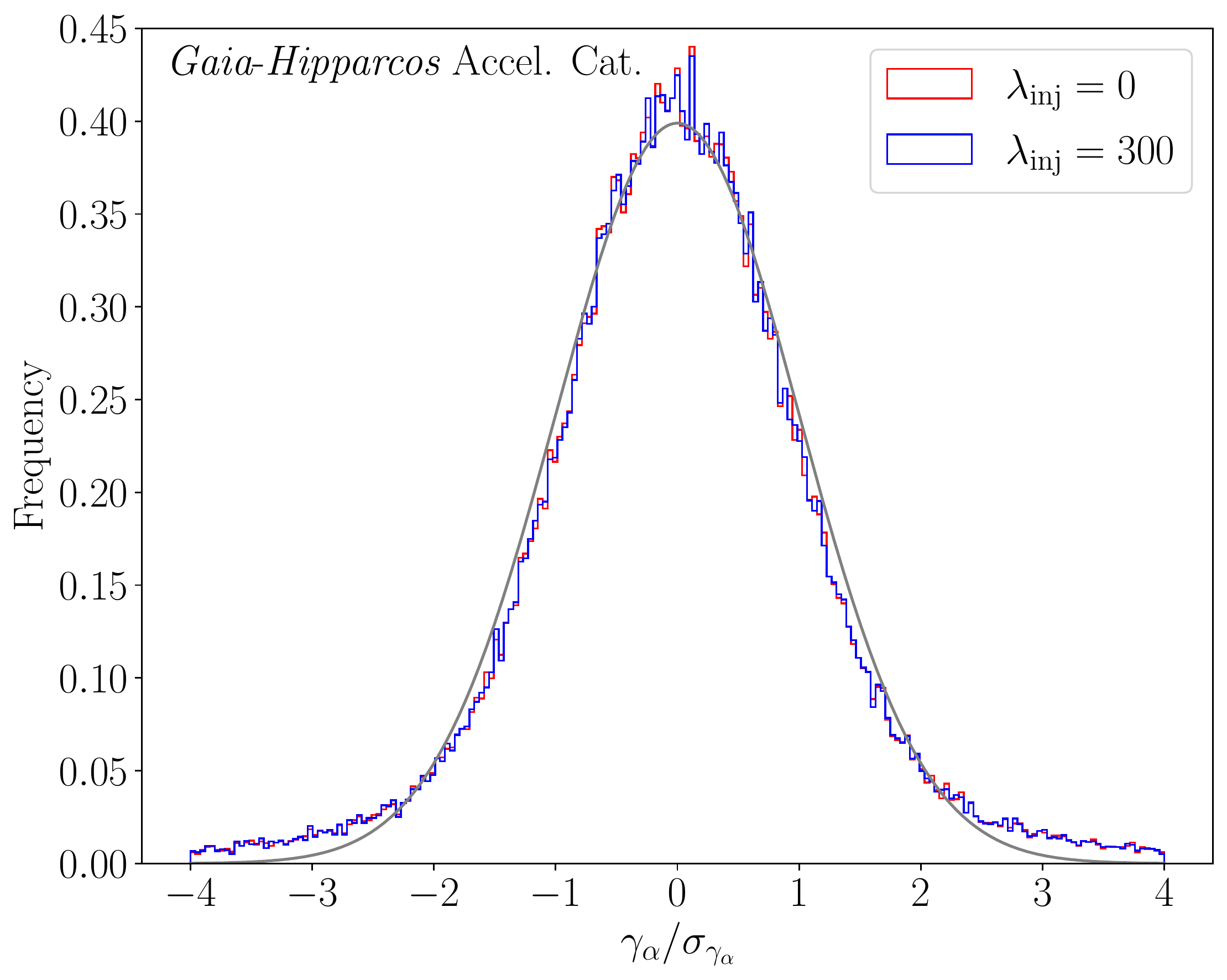}
\includegraphics[width = .49\textwidth]{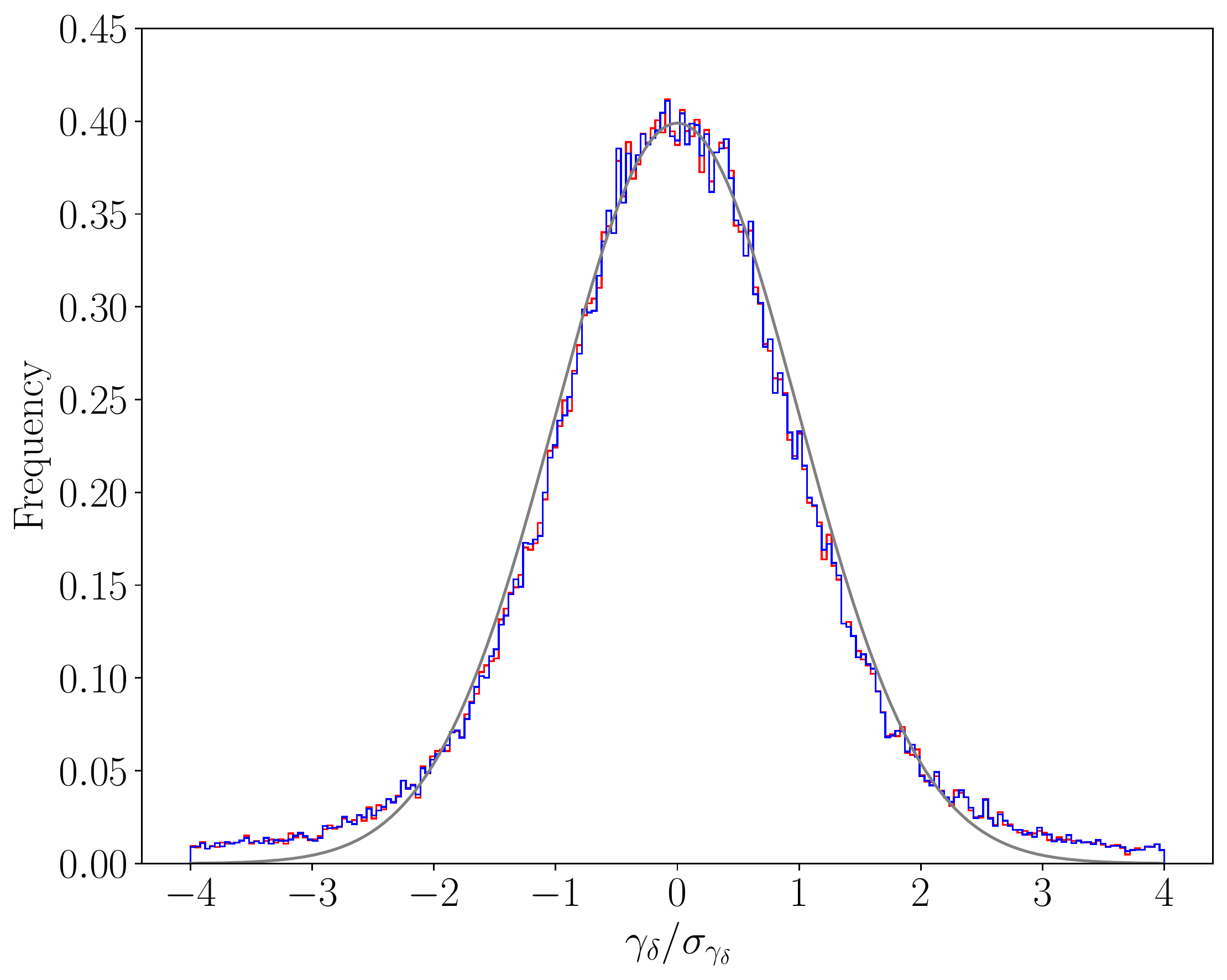}
\end{center}
\caption{As in Fig.~\ref{fig:GH_data} for the RA accelerations (left) and DEC accelerations (right) but with the inclusion of the injected signal, as indicated. Even though the injected signal is recovered at $\sim$5$\sigma$ in the joint likelihood analysis, the difference between the original data set and the hybrid data set with the injected signal is not visible at the per-star level or even at the level of the histogram of accelerations, as illustrated here. } 
\label{fig:GH_data_inj}
\end{figure}

In Fig.~\ref{fig:GH_data_inj} we histogram the RA accelerations before and after the synthetic signal injection. Interestingly, we see that despite the large value of $\lambda_{\rm inj}$ there is virtually no difference, by eye, in the histogram of the $\gamma_\alpha / \sigma_{\gamma_\alpha}$, with the same being true for the DEC accelerations. Similarly, adding in this acceleration boost has a minor change in the stars that pass our quality cuts. Before adding in the acceleration boosts we have $N = 86,201$ while after the boosts are added this number grows by two to $N = 86,203$. However, in the context of the joint likelihood the injected synthetic signal is abundantly clear, as illustrated in Fig.~\ref{fig:GH_chi2} where we show the result of the analysis (in dashed black) on the hybrid data set that includes the synthetic signal. We now find approximately 5$\sigma$ evidence for non-zero $\lambda$, and we reconstruct the correct value of $\lambda$ to within $\sim$1$\sigma$. 

\subsection{Systematic uncertainties}

\begin{figure}[htb]
\begin{center}
\includegraphics[width = .49\textwidth]{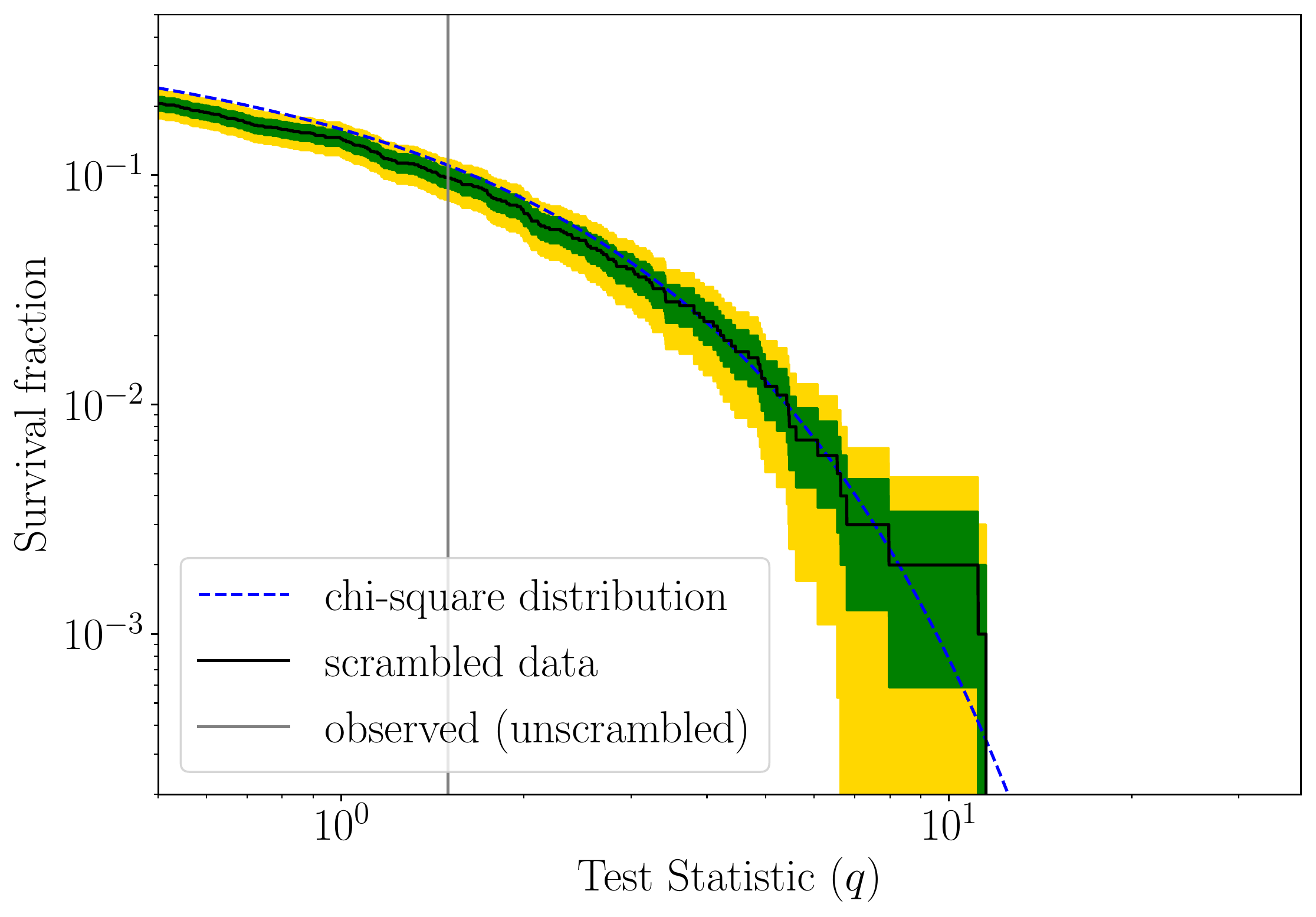}
\includegraphics[width = .49\textwidth]{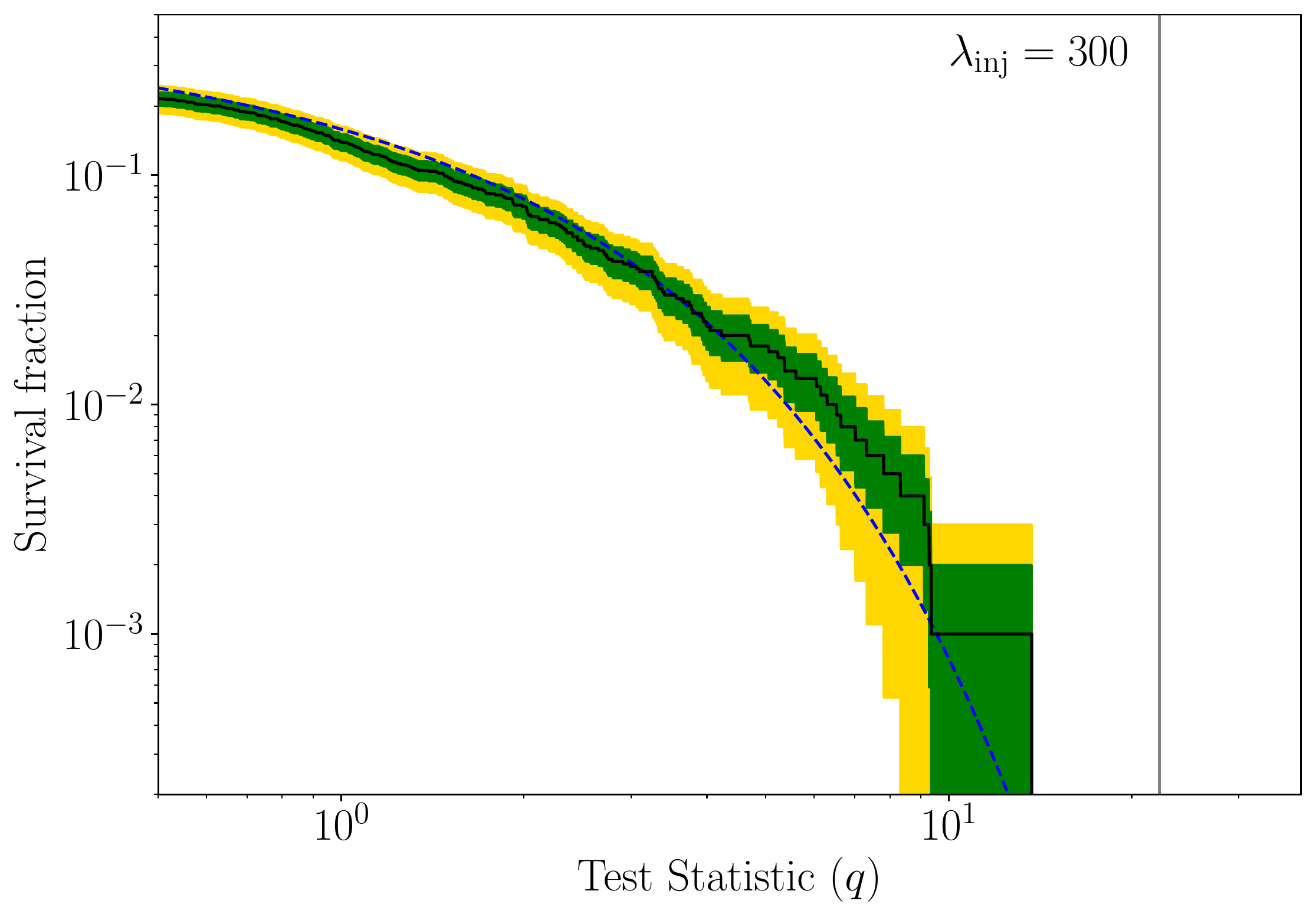}
\end{center}
\caption{ (Left) The survival fraction of test statistics $q$ over the ensemble of $10^3$ data sets generated by randomly scrambling the predictions between the model predictions and the associated stars in the {\it Gaia}-{\it Hipparcos} data (see text for details). The survival fraction shows the fraction of data sets with a $q$ larger than the value indicated on the $x$-axis. The green and gold bands indicated the statistical uncertainties arising from the finite number ($10^3$) of realizations. Under the null hypothesis we expect the survival fraction to follow the one-sided chi-square distribution with one degree of freedom, which is indicated. The observed $q$ in the unscrambled data is indicated by the vertical line. (Right) As in the left panel, but with the addition of the injected signal with $\lambda_{\rm inj} = 300$. The scrambling process largely removes the knowledge of the injected signal, as evidenced by the large gap in $q$ between the results of the unscrambled analysis (with the recovered TS shown as a vertical line) and the distribution of scrambled analyses.} 
\label{fig:GH_survival}
\end{figure}

Sources of systematic uncertainty may cause deviations of the test statistic $q$ away for the expected one-sided chi-square distribution under the null hypothesis. For example, we have assumed normally-distributed uncertainties but, as shown in Fig.~\ref{fig:GH_data}, there is evidence for non-Gaussian tails to the {\it Gaia}-{\it Hipparcos} acceleration data that are not completely removed by our quality cuts. One method to assess the impact of these systematics is to scramble the mapping from model predictions to stars. Given the large number of stars in the sample this allows us to effectively generate large numbers of null-hypothesis data sets from the real data, in the sense that even if a signal were present in the real data the signal would not be manifest in the scrambled data. However, many sources of systematic effects, such as non-Gaussianities, are not affected by the scrambling processes and thus may be probed by this test. We construct and analyze $10^3$ scrambled {\it Gaia}-{\it Hipparcos} data sets; the survival fraction for the distribution of discovery test statistics from this ensemble is illustrated in the left panel of Fig.~\ref{fig:GH_survival}, as compared to the survival fraction for the one-sided chi-square distribution expected for the null hypothesis from statistical uncertainties alone. Note that we illustrate 1$\sigma$ and 2$\sigma$ uncertainties in green and gold, respectively, on the survival fraction from the counting uncertainties associated with the finite number of scrambled data sets. The observed survival fraction matches that of the chi-square distribution to within the precision probed, indicating that systematic effects are likely subdominant compared to statistical uncertainties for this analysis. As a cross-check that this test is not affected by the inclusion of a signal, in the right panel we repeat this test on the hybrid data set constructed by injecting a synthetic signal with $\lambda_{\rm inj} = 300$ into the actual data. Importantly, the knowledge of the signal is mostly lost when the data is scrambled; none of the 1000 scrambled data sets have a test statistic nearly as large as the $q \approx 22$ observed on the unscrambled data set. There is mild evidence for an increase in the number of $q$ near $\sim$10, though even this increase is not significant within the statistical uncertainties.

The ability to test the null hypothesis by scrambling the model predictions is a unique aspect of the stellar acceleration approach to measuring the Galactic potential that will enable future studies to address systematic uncertainties in a data-driven approach, for example by the use of ``spurious-signal" nuisance parameters that may account for the mismodeling, if necessary (see, {\it e.g.},~\cite{Aad:2014eha}). 

\section{Extended results for {\it Gaia} projections}
\label{app:extended}

\begin{figure}[htb]
\begin{center}
\includegraphics[width = .49\textwidth]{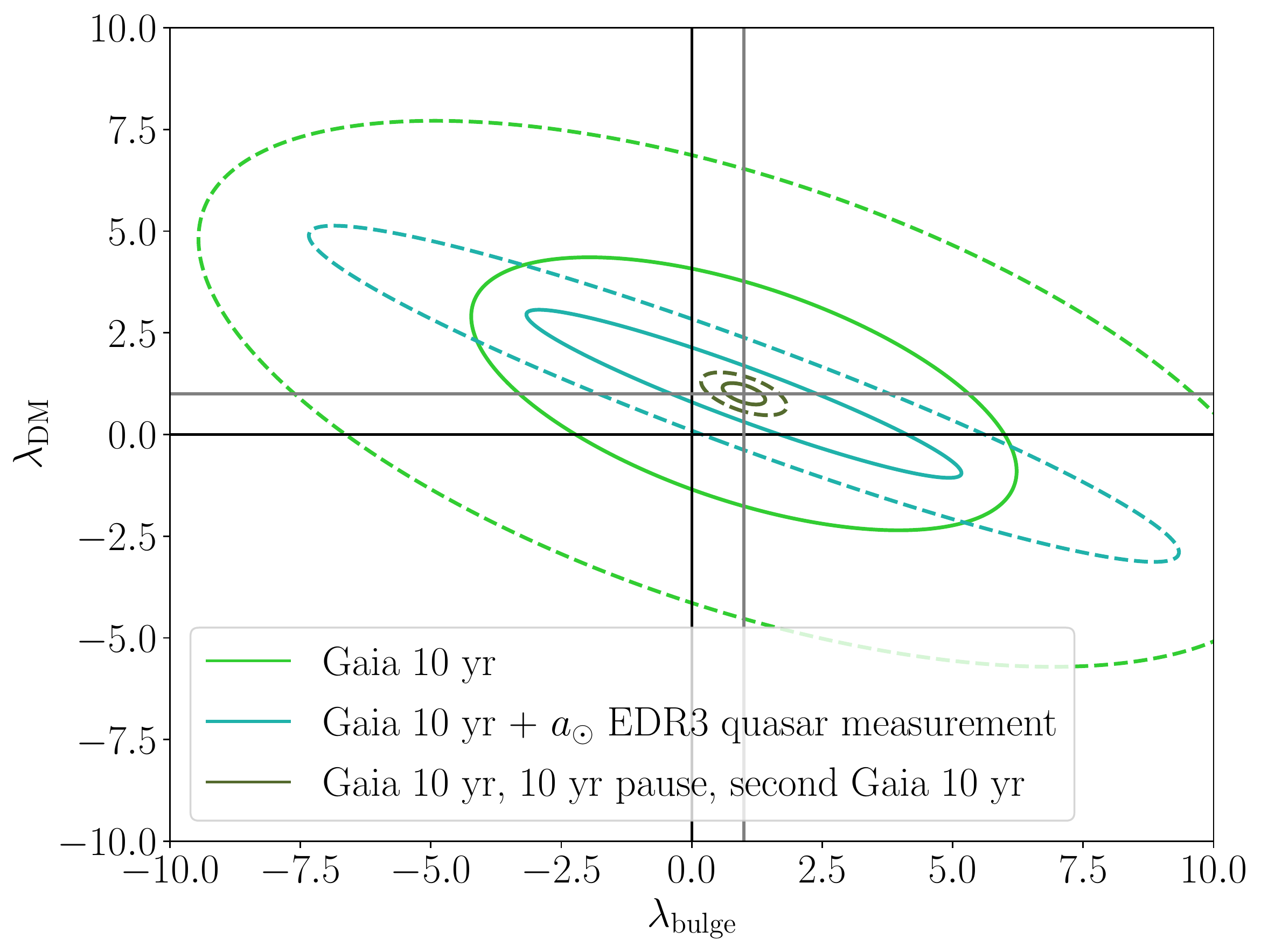}
\includegraphics[width = .49\textwidth]{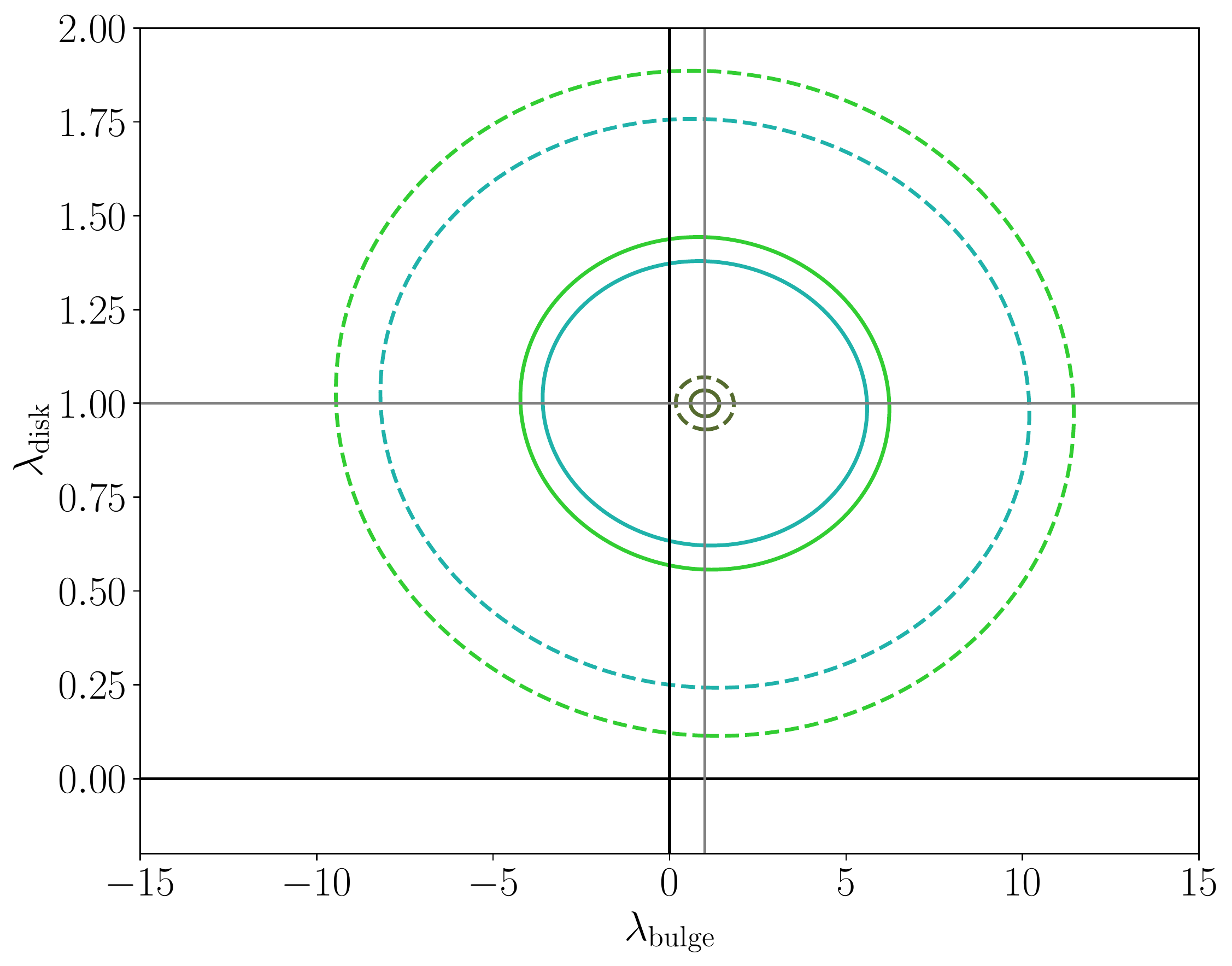}
\end{center}
\caption{As in Fig.~\ref{fig:gaia_ill} but for the $\lambda_{\rm disk}$-$\lambda_{\rm bulge}$ and $\lambda_{\rm DM}$-$\lambda_{\rm bulge}$ covariance projections.
} 
\label{fig:gaia_ill_bulge}
\end{figure}

In Fig.~\ref{fig:gaia_ill} we showed the projected covariance between $\lambda_{\rm disk}$ and $\lambda_{\rm DM}$ for astrometric data sets. In Fig.~\ref{fig:gaia_ill_bulge} we illustrate the corresponding figures showing the $\lambda_{\rm disk}$-$\lambda_{\rm bulge}$ and $\lambda_{\rm DM}$-$\lambda_{\rm bulge}$ covariance projections. As evident already from the illustration in Fig.~\ref{fig:accel_ill}, the bulge is harder to constrain than the DM and disk components, though future data sets may also be sensitive to this component.

Note that our {\it Gaia} projections assume fiducial disk, bulge, and DM models. It is possible that the real density distributions are different than assumed here, which could affect our projections. For example, we normalized the DM density distribution to the local density $\rho_\odot = 0.008 M_\odot / {\rm pc}^3$, but one of the main purposes of the proposed analysis framework is to determine this value directly. If $\rho_\odot$ is at a different value then the expected detection significance for the DM halo would scale linearly with $\rho_\odot$. We also assume in this work a fiducial scale radius of $r_s = 16$ kpc. As discussed in the main {\it Letter}, future surveys may have the ability to measure this scale radius precisely, though it has a subdominant effect of the detection significance of the DM component. The morphological parameters of the disk may also be constrained with future data sets using the acceleration approach. In that context it is important that future work investigate, for example, the effect that mismodeling the disk component could have on determinations of the DM component.

Lastly, we note that {\it Gaia} may also measure radial accelerations by looking for changes in the radial velocities of the stars using the on-board Radial Velocity Spectrometer instrument. However, we estimate that the radial acceleration measurements from {\it Gaia} would be less constraining than the angular accelerations discussed here. Still, it is worth keeping in mind that future surveys may improve their radial velocity sensitivity to the point that the inclusion of this data within the context of the joint likelihood would be beneficial.

\end{document}